\documentclass[letterpaper]{article} 
\usepackage{aaai25}  
\usepackage{times}  
\usepackage{helvet}  
\usepackage{courier}  
\usepackage[hyphens]{url}  
\usepackage{graphicx} 
\usepackage{natbib}  
\usepackage{caption} 
\usepackage{algorithm}
\usepackage{algorithmic}

\usepackage{amsmath}
\usepackage{cleveref}

\usepackage{newfloat}
\usepackage{listings}
\DeclareCaptionStyle{ruled}{labelfont=normalfont,labelsep=colon,strut=off} 
\floatstyle{ruled}
\newfloat{listing}{tb}{lst}{}
\floatname{listing}{Listing}
\title{Advancing Loss Functions in Recommender Systems: A Comparative Study with a Rényi Divergence-Based Solution}
\author {
    Shengjia Zhang\textsuperscript{\rm{1,2}}, 
    Jiawei Chen\textsuperscript{\rm{1,2,3,}}\thanks{Corresponding author: sleepyhunt@zju.edu.cn}, 
    Changdong Li\textsuperscript{\rm{2}}, 
    Sheng Zhou\textsuperscript{\rm{2}}, 
    Qihao Shi\textsuperscript{\rm{2}}, \\
    Yan Feng\textsuperscript{\rm{1,2}}, 
    Chun Chen\textsuperscript{\rm{1,2}}, 
    Can Wang\textsuperscript{\rm{1,3}}
}
\affiliations {
    \textsuperscript{\rm{1}} State Key Laboratory of Blockchain and Data Security, Zhejiang University\\
    \textsuperscript{\rm 2} College of Computer Science, Zhejiang University, China\\
    \textsuperscript{\rm 3} Hangzhou High-Tech Zone (Binjiang) Institute of Blockchain and Data Security\\
\{shengjia.zhang, sleepyhunt, lichangdongtw, zhousheng$\_$zju, shiqihao321, fengyan, chenc, wcan\}@zju.edu.cn
}

\usepackage{bibentry}

\newcommand{\eg}{\emph{e.g., }}
\newcommand{\lossName}{\text{DrRL}}
\newcommand{\fullLossName}{\textbf{\underline{D}istributional \underline{R}obust \underline{R}ényi \underline{L}oss}}

\newtheorem{lemma}{Lemma}

\newcommand{\userSet}{\mathcal{U}}
\newcommand{\itemSet}{\mathcal{I}}

\newcommand{\uiPair}{(u,i)}

\newcommand{\obSet}{\mathcal{O}}
\newcommand{\posU}{\mathcal{I}_{u}^+}
\newcommand{\negU}{\mathcal{I}_{u}^-}

\newcommand{\E}{\mathbb{E}}

\newcommand{\excetra}{\text{et al.}}
\newcommand{\optimalBeta}{\beta_u}
\newcommand{\normalBeta}{\beta}

\newcommand{\loss}{\mathcal{L}}
\newcommand{\similarity}{f}
\newcommand{\wStar}{\gamma^{*}}
\newcommand{\wNormal}{\gamma}
\newcommand{\negFac}{c_{\gamma}(\eta)}

\newcommand{\distFamily}{Q}

\newcommand{\SLNegWeight}{w_{uj}}
\newcommand{\cclWeight}{\alpha}

\newcommand{\ie}{\emph{i.e., }}

\newcommand{\etc}{etc}
\newcommand{\st}{\textrm{s.t. }}

\newcommand{\confer}{\textit{cf. }}

\newcommand{\userNegDis}{P_u^-}

\newcommand{\userUncertainDis}{Q_u}
\DeclareMathOperator*{\esssup}{ess\,sup}

\usepackage{multirow}
\usepackage{multicol}
\usepackage{booktabs}
\usepackage{amssymb}
\usepackage{subfigure}
\usepackage{bigstrut}       
\usepackage{makecell}       

\usepackage{colortbl}

\usepackage{tabularx}       
\usepackage{booktabs}

\begin{document}

\maketitle

\begin{abstract}
Loss functions play a pivotal role in optimizing recommendation models. Among various loss functions, Softmax Loss (SL) and Cosine Contrastive Loss (CCL) are particularly effective. Their theoretical connections and differences warrant in-depth exploration. This work conducts comprehensive analyses of these losses, yielding significant insights: 1) Common strengths --- both can be viewed as augmentations of traditional losses with Distributional Robust Optimization (DRO), enhancing robustness to distributional shifts; 2) Respective limitations --- stemming from their use of different distribution distance metrics in DRO optimization, SL exhibits high sensitivity to false negative instances, whereas CCL suffers from low data utilization. To address these limitations, this work proposes a new loss function, DrRL, which generalizes SL and CCL by leveraging Rényi-divergence in DRO optimization. DrRL incorporates the advantageous structures of both SL and CCL, and can be demonstrated to effectively mitigate their limitations. Extensive experiments have been conducted to validate the superiority of DrRL on both recommendation accuracy and robustness.
\end{abstract}
\begin{links}
\link{Code}{https://github.com/cynthia-shengjia/AAAI-2025-DrRL
}
\end{links}

\section{Introduction}

Recommender Systems (RS)~\cite{ricci2021recommender,gao2023cirs,cui2024distillation,liao2024llara} are pivotal in delivering personalized suggestions across various online services. Collaborative Filtering (CF) has emerged as an effective method, learning user preferences from historical interactions~\cite{he2017neural,shi2014collaborative}. Loss functions, which direct the optimization pathways of models, are critically important. Traditionally, RS have primarily utilized point-wise losses, such as Binary Cross-Entropy (BCE) \cite{johnson2014logistic} and Mean Squared Error (MSE) \cite{pan2008one}, or pairwise losses like Bayesian Personalized Ranking (BPR) \cite{rendle2009bpr}. Recent years have witnessed a booming interest in exploring novel loss functions, leading to substantial research advancements in this domain.

\begin{figure}[t]
  \centering
  \includegraphics[width=\linewidth]{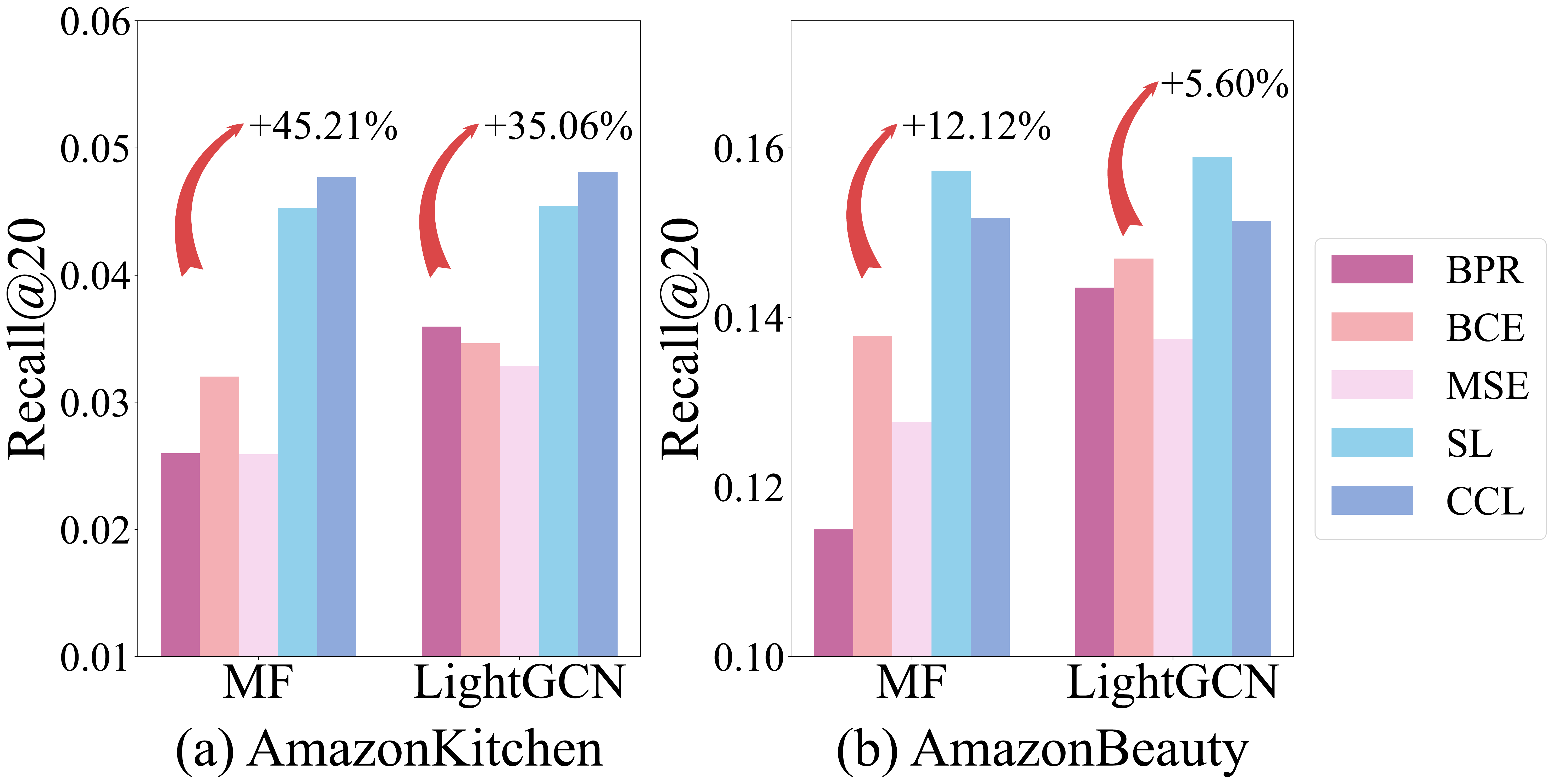}
  \caption{Illustration of how SL and CCL outperforms traditional losses on two recommendation backbones. The detailed experimental settings refer to section \ref{sec:experiment}.}
    \label{fig1:the intro comparison with BCE}
\end{figure}

By scrutinizing various existing loss functions, we identify two particularly effective types: 1) Softmax Loss (SL) \cite{wu2024effectiveness}, which employs a softmax function to normalize model predictions and enhance the scores of positive instances relative to negatives; 2) Cosine Contrastive Loss (CCL) \cite{mao2021simplex}, which augments traditional losses by integrating a truncation mechanism that filters out negative instances with low prediction scores. Figure~\ref{fig1:the intro comparison with BCE} provides empirical evidence ---  SL and CCL outperform traditional losses by a significant margin (over 20\% on average) on two real-world datasets.  These impressive observations pose a compelling question: \textbf{What are the inherent common strengths of SL and CCL, and can we develop a new loss function based on these advantages?}

This study conducts comprehensive theoretical analyses to demystify this aspect.  Despite their distinct structures, we demonstrate that both SL and CCL can equivalently augment traditional losses with \textit{Distributional Robust Optimization} (DRO) \cite{lin2022distributionally}.  DRO is a theoretically robust optimization framework that extends model optimization beyond observed training distributions to a broader family of potential distributions with perturbations. Given the common occurrence of distribution shifts in RS --- such as evolving user preferences~\cite{wang2022causal} and inherent biases in data collection~\cite{chen2020bias} --- the efficacy of DRO-enhanced losses is anticipated.

Beyond their shared advantages, the distinctions and limitations of SL and CCL are also revealed. The primary difference lies in the types of distribution perturbations employed in optimizations --- SL uses perturbations constrained by KL-divergence~\cite{wu2024understanding}, whereas CCL uses perturbations constrained by worst-case regret-divergence~\cite{duchi2021learning}. These choices lead to the following limitations:

\begin{itemize}
    \item \textbf{SL’s sensitivity to noise}:  SL employs a KL-divergence-based DRO, this type of DRO has been shown to be highly sensitive to noise ~\cite{zhai2021doro,nietert2024outlier} (\eg false negative instances in RS). Specifically, SL assigns disproportionately large weights to negative instances with higher scores, governed by an exponential function. False negatives, which are common in RS and often result from user unawareness rather than disinterest \cite{chen2019samwalker,gao2022self,chen2023adap}, receive excessive emphasis, leading to performance degradation.
    \item \textbf{CCL’s low data utilization}: While the truncation mechanism promises enhanced out-of-distribution robustness and model convergence, it naturally reduces data utilization, which is particularly serious in CCL. Specifically, we find that an excessively large proportion (often over 90\%) of negative instances are filtered out by CCL. Although these lower-scored negative instances individually contribute less to the gradient than higher-scored ones, their large quantity could still offer valuable training signals. Additionally, the optimal truncation threshold is treated as a hyper-parameter requiring manual tuning, which incurs additional parameter tuning efforts.
\end{itemize}

To address these limitations, we propose a new loss function, named \fullLossName \ ($\lossName$), which generalizes SL and CCL by employing Rényi-divergence-constrained DRO optimization. Rényi divergence~\cite{van2014renyi}, a broad family of divergences that includes common forms like KL-divergence and $\chi^2$ divergence, allows our $\lossName$ to inherit and enhance the strengths of SL and CCL while circumventing their drawbacks: 1) $\lossName$ retains the weighting strategy from SL but provides flexible control over the shape of the weighting distribution, leveraging a polynomial distribution whose order can be flexibly adjusted. This mitigates the excessive impact of false negatives, thereby enhancing robustness to noise. 2) $\lossName$ also inherits the truncation strategy from CCL but demonstrates better data utilization. Additionally, $\lossName$ offers a theoretically sound strategy to learn fine-grained truncation thresholds, avoiding the need for tedious parameter tuning.

In summary, this work makes the following contributions:

\begin{itemize}
    \item Conducting comprehensive theoretical and empirical analyses to elucidate the theoretical connections between SL and CCL and highlight their limitations.
    \item Proposing a new recommendation loss function, Distributional Robust Rényi Loss ($\lossName$), which leverages Rényi-divergence-constrained DRO to inherit the advantages and circumvent the limitations of SL and CCL.
    \item Performing extensive experiments to demonstrate the effectiveness and robustness of $\lossName$ over existing losses.
\end{itemize}

\section{Preliminaries}

\subsection{Task Formulation}

This work focuses on collaborative filtering (CF) \cite{he2017neural,shi2014collaborative}, a conventional recommendation scenario. Let $\userSet$ and $\itemSet$ denote a user set and an item set. Let $\obSet \subset \userSet \times \itemSet$ denote the set of the observed interactions. A user-item pair $\uiPair$ in $\obSet$ indicates that the user $u$ has interacted with the item $i$ (\eg click, purchase, etc). For convenience, we define the set of items that user $u$ has interacted with as the positive item set $\posU = \{i \in \itemSet: \uiPair \in \obSet\}$, with the remaining items forming the negative item set $\negU = \itemSet \setminus \posU$. The goal of RS is to recommend items to each user that they may be interested in. 

Embedding-based methods are widely utilized in RS. These methods first map users and items into $d$-dimensional embeddings $\mathbf{e}_u$ and $ \mathbf{e}_i$, and then generate model predictions $f(u,i)$ with embedding similarity.  Recent work has demonstrated cosine similarity is particularly effective~\cite{chen2023adap}, \ie  $f(u,i)=\frac{\mathbf{e}_u^{\top} \mathbf{e}_i}{\Vert \mathbf{e}_u\Vert \Vert \mathbf{e}_i \Vert}$. For convenience, this work adopts cosine similarity for analysis, though our findings can be generalized to other similarity metrics.

\subsection{Recommendation Loss Functions}

Beyond the traditional point-wise and pair-wise loss functions, recent years have witnessed a surge of research on novel recommendation losses. Two representative types are:

\textbf{Softmax Loss (SL)~\cite{wu2024effectiveness}.} SL normalizes model predictions using the softmax function, enhancing the scores of positive instances relative to negative ones. The loss for each user $u$ can be written as:
\begin{equation}
\label{eqs:SL empirical}
\begin{aligned}
&\loss_{\text{SL}}(u) =
-\frac{1}{\vert \posU \vert}\sum\limits_{i \in \posU} \log\left(\frac{\exp(\similarity(u,i) / \tau)}{\sum_{j \in \negU }\exp(\similarity(u,j) / \tau)} \right) \\
=&-\frac{1}{\vert \posU \vert} \sum\limits_{i \in \posU} \similarity(u,i)/\tau + \log\left(\sum_{j \in \negU }\exp(\similarity(u,j) / \tau)\right),
\end{aligned}
\end{equation}
where $\tau$ is a temperature parameter for rescaling the model predictions. Building on SL, recent works have proposed various improvements, such as enhancing popularity debiasing~\cite{zhang2022incorporating} and out-of-distribution (OOD) robustness~\cite{wu2024bsl}. Since SL forms the foundation for these variants and our analysis can be generalized to them, we focus on the basic SL for our analysis.

\textbf{Cosine Contrastive Loss (CCL)~\cite{mao2021simplex}.} CCL incorporates a \textit{truncation} mechanism in the classical point-wise loss:
\begin{equation}
\label{eqs:CCL empirical}
    \loss_{\text{CCL}}(u) = -\frac{1}{\vert \posU \vert} \sum\limits_{i \in \posU}\similarity(u,i) + \frac{\cclWeight}{\vert \negU \vert}\sum_{j \in \negU} (\similarity(u,j) - \beta)_+,
\end{equation}
where the symbol $(\cdot)_+$ denotes $\max\{\cdot,0\}$, $\beta$ is a margin parameter controlling the truncation threshold, and $\cclWeight$ is a parameter that rescales the contribution of the negative part. The truncation mechanism in CCL filters out negative instances with scores lower than $\beta$ during the model training.

As can be observed, SL and CCL have quite different loss structures, particularly in their treatment of the negative part. In the following sections, we will explore their connections and common strengths.

\subsection{Distributionally Robust Optimization}
The success of machine learning models typically relies on the assumption of independent and identically distributed (IID), \ie test data is sampled from the same distribution as the observed training data. However, this assumption often fails in real-world scenarios, leading to performance degradation.  \textit{Distributionally Robust Optimization} (DRO) has been demonstrated to be effective in mitigating this issue~\cite{lin2022distributionally}. DRO extends model optimization beyond the observed training distribution to a broader family of potential distributions with perturbations. Specifically, DRO aims to minimize the worst-case expected loss over a set of potential distributions $Q$, which surround the observed training distribution $Q_0$ and are constrained by a distance metric $D(Q, Q_0)$ within a radius $\eta$. The objective is formulated as follows:
\begin{equation}
    \begin{aligned}
       & \loss_{DRO}= \max\limits_{\distFamily}\E_{x \sim \distFamily}[\loss(x;\theta)] \quad \st D({Q},{Q}_0) \leq \eta 
    \end{aligned}
\end{equation}
where models are optimized under the potential distributions $Q$, which can be understood as an ``adversary'', empowering model robustness with adversarial distributional perturbations.

\section{Analyses on CCL and SL}
In this section, we first conduct theoretical analyses to reveal the common strengths of SL and CCL, followed by a discussion of the inherent limitations in these loss functions.

\subsection{Connections between CCL and SL}
Given the effectiveness of both CCL and SL, uncovering their shared strengths is valuable, as it not only deepens our understanding of loss function mechanisms, but also inspires the development of new loss functions. Recent work~\cite{wu2024understanding} has shown that SL can be equivalent to performing DRO on the negative item distribution. In this work, we further demonstrate that CCL also exhibits this advantageous property, despite its loss structure being quite different from SL.

Specifically, let $\userNegDis$ represent the observed negative item distribution of user $u$, \ie the uniform distribution over the negative item set $\mathcal I_u^-$. We present the following lemma: 
\begin{lemma}
  \label{lem: CCL and SL DRO}
   Optimizing recommendation models with SL and CCL can both be equivalent to solving the following DRO objective:
      \begin{equation}
      \label{eqs:Distributionally Contrastive Loss}
          \loss_{\text{DRO}}(u) = -\frac{1}{\vert \posU \vert}\sum\limits_{i \in \posU}\similarity(u,i) + \max\limits_{\userUncertainDis}\E_{j \sim \distFamily_u}\left[ \similarity(u,j)\right].
      \end{equation}
SL is constrained by KL-divergence within robust radius $\eta$:
  \begin{equation}
  \label{eqs:SL set}
        D_{\text{KL}}(\userUncertainDis,\userNegDis) := \sum\limits_{j \in \negU} \userUncertainDis(j) \log\frac{\userUncertainDis(j)}{\userNegDis(j)} \le \eta,
  \end{equation}
while CCL is constrained by the worst-case regret-divergence within robust radius $\log(\alpha)$:
\begin{equation}
\label{eqs:CCL set}
   D_{\text{WR}}(\userUncertainDis, \userNegDis) := \sup\limits_{j \in \negU}\log  \frac{\userUncertainDis(j)}{\userNegDis(j)}  \le \log(\cclWeight),
\end{equation}
where $\cclWeight$ is the rescale parameter in Eq.(\ref{eqs:CCL empirical}).
\end{lemma}
The proof is given in \Cref{appendix-CCL-and-SL}. This lemma clearly demonstrates the common strengths of SL and CCL. Both can be understood as DRO-enhanced versions of the classical point-wise objective: 
\begin{equation}
    \loss_{basic}(u) = -\frac{1}{\vert \posU \vert}\sum\limits_{i \in \posU}\similarity(u,i) + \E_{j \sim P^-_u}\left[ \similarity(u,j)\right].
\end{equation}
SL and CCL improve $\loss_{\text{basic}}(u)$ by leveraging DRO on negative side, where the model is optimized on a set of potential distributions $Q_u$ with distributional perturbations. Due to this adversarial optimization mechanism, DRO endows SL and CCL with robustness against distribution shifts. It is particularly noteworthy that distribution shifts are ubiquitous in RS, \eg user preferences typically evolve over time~\cite{wang2022causal}, and the training data is often polluted by various biases~\cite{chen2020bias,chen2021autodebias,gao2023alleviating}. This property makes SL and CCL particularly effective.
\begin{figure}[t]
  \centering
  \includegraphics[width=0.95\linewidth]{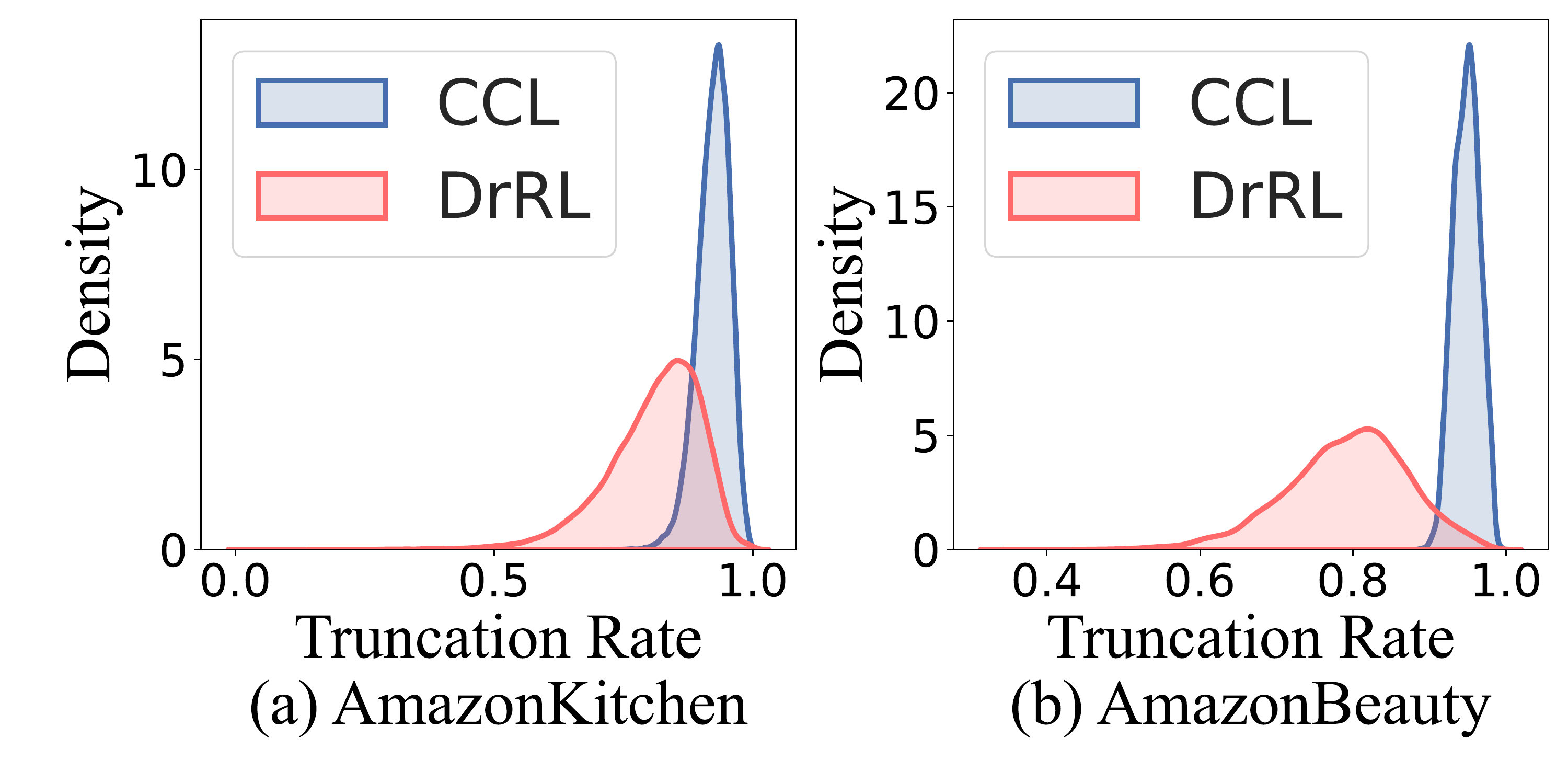}
  \caption{The truncation ratios of $\lossName$ and CCL over users in two real-world datasets.}
    \label{fig:poor data utilization of CCL}
\end{figure}

\subsection{Limitations of CCL and SL}
The above analyses reveal that both CCL and SL can be considered as DRO-enhanced losses, but with different divergence constraints. These different choices of divergences lead to different limitations:
  
\textbf{SL is highly sensitive to noise.} Recent work~\cite{nietert2024outlier} has demonstrated that DRO with KL-divergence is highly sensitive to noise.  To understand this effect, we can derive the worst-case distribution, $\ie$$Q^{*}_u = \mathop{\arg\max}\limits_{\userUncertainDis} \E_{j \sim \userUncertainDis}\left[\similarity(u,j)\right]$, under which the model is finally optimized:
\begin{equation}
\begin{aligned}
    Q^{*}_u(j) =&  \SLNegWeight \userNegDis(j) \\
    \SLNegWeight =& \frac{\exp(\similarity(u,j) / \tau)}{\mathbb{E}_{j \sim \userNegDis} \left[\exp( \similarity(u,j) / \tau )\right]} \propto \exp(\similarity(u,j)/\tau)
\end{aligned}
\end{equation}
The proof is given by Wu~\excetra~(see Appendix A.6 in ~\cite{wu2024understanding}). This implies that SL assigns a weight $\SLNegWeight$ to each negative instance, with the weights $w_{uj}$ proportional to the exponential of the prediction scores. Given the explosive nature of the exponential function and the fact that $\tau$ is usually set to a relatively small value (\eg $\tau=0.2$), the weight distribution becomes highly skewed. This skew leads to negative instances with higher scores exerting a disproportionate influence on model training. Such a characteristic renders SL particularly vulnerable to noise, such as false negative instances, which are prevalent in RS. Given the vast number of items and the limited attention of users, some negative instances typically result merely from users' lack of awareness of these items, rather than an active dislike \cite{chen2019samwalker,gao2022self,wang2024llm4dsr}. These false negative items, potentially sharing similar features with positive items, can easily receive large prediction scores $f(u,i)$. Consequently, they attain excessively large weights in model training, potentially dominating the optimization directions and severely degrading performance. Table~\ref{table:noise to neg of SL} presents the empirical evidence, demonstrating that noisy data receive significantly large weights in SL, typically over 20 times.

Interestingly, recent work~\cite{wu2024bsl} claims that SL exhibits robustness to noisy data through DRO. However, we argue that this is not the case. While DRO can indeed enhance model robustness to distribution shifts, it can also increase noise sensitivity rather than decrease, as demonstrated by various studies on DRO~\cite{zhai2021doro,nietert2024outlier}. Our analysis also shows that noisy data in SL contribute more, rather than less, to the optimization process. This is further evidenced by empirical results from experiments involving false negative instances (\confer Figure 8 in~\cite{wu2024bsl}), where the improvements of SL over other baselines do not significantly increase and sometimes even decline as the noise ratio increases.

One might suggest adjusting $\tau$ to improve noise resistance. However, $\tau$ is a dual parameter of the robust radius $\eta$. As discussed in~\cite{wu2024understanding}, increasing $\tau$ would naturally decrease the robust radius $\eta$, thereby reducing the model's robustness to distribution shifts. Adjusting $\tau$ would significantly decline the merit of SL. 
  
\textbf{CCL suffers from low data utilization.}  The truncation mechanism employed by CCL promises to enhance model OOD robustness and accelerate convergence. However, it also naturally reduces data utilization, filtering out a large portion of instances during training, which is particularly severe in CCL. Figure~\ref{fig:poor data utilization of CCL} illustrates the ratio of filtered items across users with the optimal $\beta$. The ratio is quite extreme, with over 90\% of negative instances often being filtered out. Although these lower-scored negative instances individually contribute less to the gradient than higher-scored ones, their large quantity provides valuable training signals.

\begin{figure}[t]
  \centering

  \hspace{-0.2in}
  \subfigure{
    \includegraphics[height=1.30in]{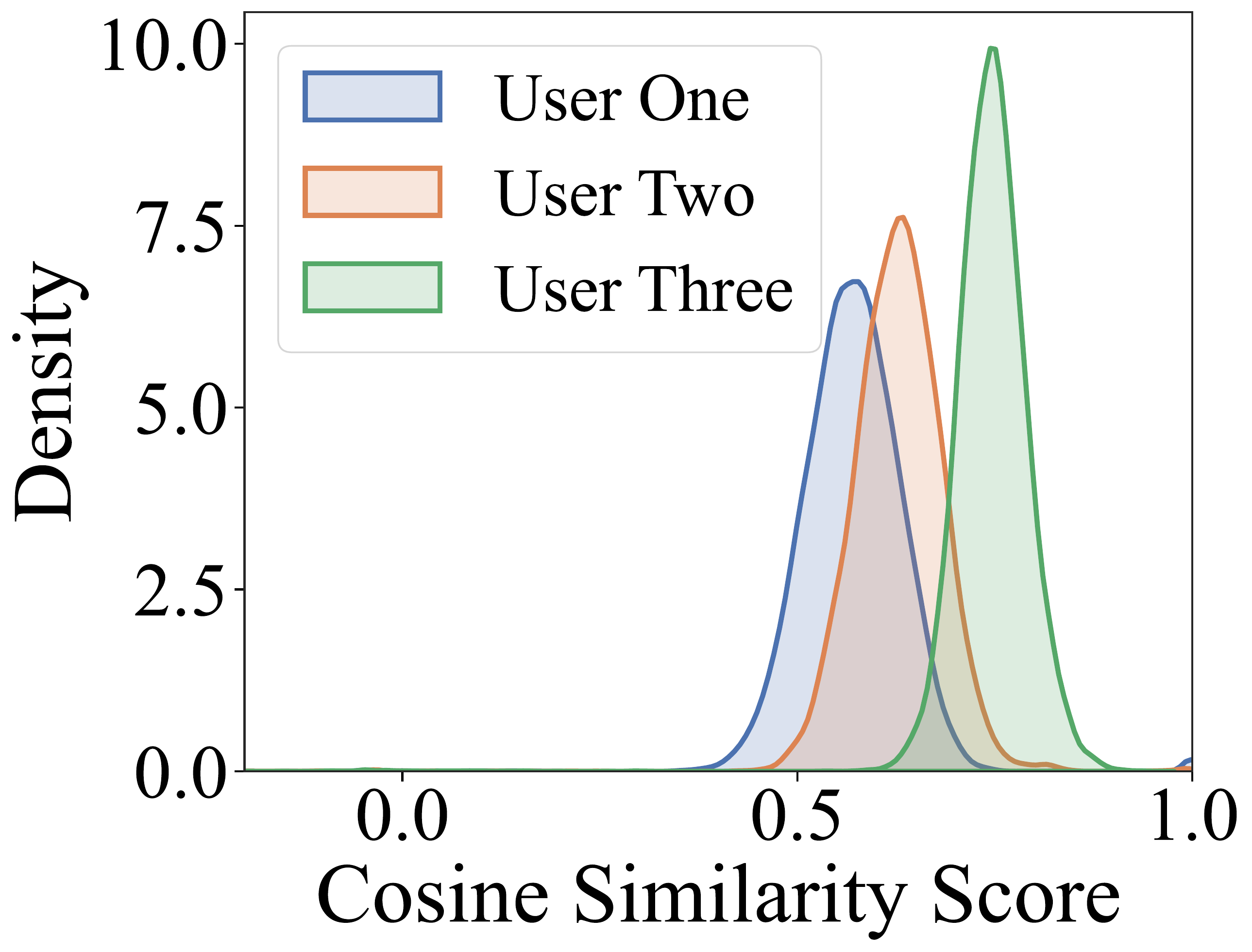}
    }
  \hspace{-0.105in}
  \subfigure{
    \includegraphics[height=1.30in]{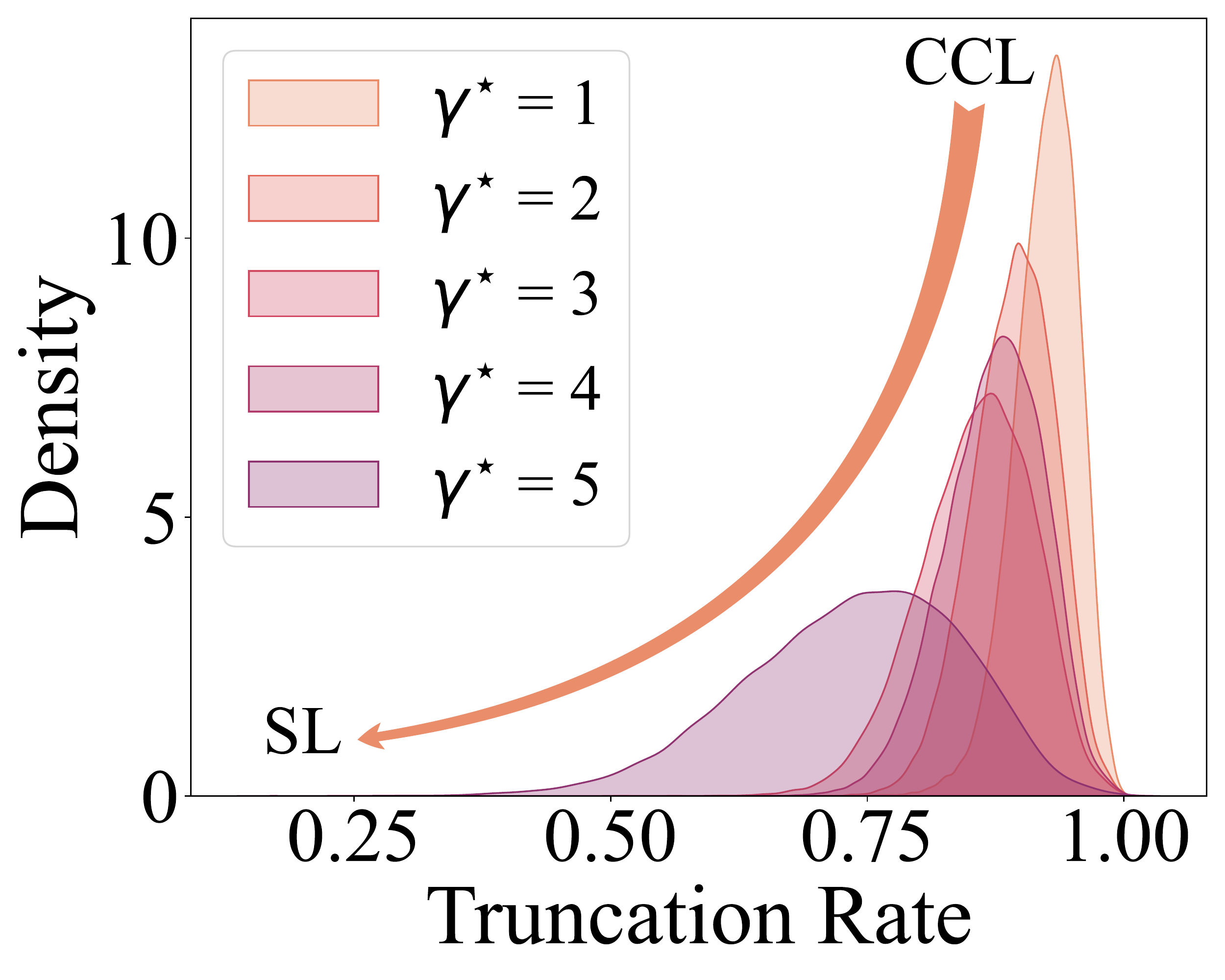}
    }

  \caption{Left: The distribution of $f(u,i)$ for three randomly sampled users; Right: the truncation rate for users with varying $\gamma$ in the dataset AmazonKitchen.} 

    \label{fig: user preference diff and cut trend}
\end{figure}

Another limitation of CCL is that it treats $\beta$ as a hyper-parameter, necessitating manual adjustments and incurring substantial tuning efforts. The reason for this is that the authors of CCL designed it heuristically, without considering its equivalence to DRO. Moreover, we argue that assigning the same $\beta$ to all users may not be optimal. Figure~\ref{fig: user preference diff and cut trend} illustrates that users have diverse preference distributions $f(u,i)$ over items. This is a common phenomenon in practice, since some users have broader preferences, while others may be more critical. Such diversity naturally motivates us to pursue a personalized $\beta$.

\section{Distributionally Robust Rényi Loss}

To further capitalize on the strengths of SL and CCL while circumventing their limitations, a straightforward approach is to maintain their DRO-enhanced objective (\confer Eq.(\ref{eqs:Distributionally Contrastive Loss})) while utilizing a more appropriate divergence measure. This work proposes leveraging Rényi divergence~\cite{van2014renyi}, a natural generalization of both KL-divergence and worst-case regret-divergence used by SL and CCL. The proposed Distributionally Robust Rényi Loss ($\lossName$) is formulated as follows:
\begin{equation}
\small{
\begin{aligned}
\label{eqs:renyi optimization objective}
        \loss_{\lossName}(u) = -&\frac{1}{\vert \posU \vert}\sum\limits_{i \in \posU} \similarity(u,i) + \max\limits_{\userUncertainDis} \E_{j \sim \userUncertainDis}\left[f(u,j)\right] \\
        \st & D_{\wNormal}(\userUncertainDis,\userNegDis) := \sum\limits_{j \in \negU} \userNegDis(j)\phi_{\wNormal}\left(\frac{\userUncertainDis(j)}{\userNegDis(j)}\right)  \le \eta \\
        & \phi_{\wNormal}(t) = \frac{1}{\wNormal(\wNormal - 1)}(t^{\wNormal} - \wNormal t + \wNormal - 1)
\end{aligned}
}
\end{equation}

Rényi divergence $D_{\wNormal}(\userUncertainDis,\userNegDis)$ offers enhanced flexibility by using a parameter $\gamma$ to adjust the polynomial relationships of the probability distance measure with the probability ratio. Here, we adopt the Cressie-Read family of Rényi divergence~\cite{duchi2021learning}, which offers analytical benefits. Notably, while Eq.(\ref{eqs:renyi optimization objective}) is complex and challenging to optimize directly, it can be substantially simplified with the following lemma:
\begin{lemma}
\label{lemma2: renyi loss}
    Suppose $\wNormal \in (1,+\infty)$, optimizing the objective in Eq.(\ref{eqs:renyi optimization objective}) is equivalent to optimizing:
    

    \begin{equation}
\label{eqs: close form renyi loss}
\small{
        \begin{aligned}
&\loss_{\lossName}(u) = -\frac{\sum\limits_{i \in \posU}\similarity(u,i)}{\vert \posU \vert}    \!+\!\negFac\!\!\left[ \!\frac{1}{\vert \negU \vert}\!\!\sum\limits_{j \in \negU} \!(\similarity(u,j)\! - \!\optimalBeta)_+^{\wStar}\!\right]^{\!\!\frac{1}{\wStar}} \\
&\st \optimalBeta = \mathop{\arg\min}\limits_{\beta} \Big\{\normalBeta + \negFac  \left[\frac{1}{\vert \negU \vert} \sum\limits_{j \in \negU}(\similarity(u,j) - \normalBeta)_+^{\wStar}\right]^{\frac{1}{\wStar}}\!\!\!\!\Big\}
        \end{aligned}
            }
    \end{equation}
where $\wStar = \frac{\wNormal}{\wNormal - 1}$, and $\negFac = (1 + \wNormal(\wNormal - 1)\eta)^{\frac{1}{\wNormal}}$.
\end{lemma}


The proof is presented in \Cref{proof: Appendix 2}. This lemma yields a simple closed-form expression of $\lossName$. It is straightforward and intuitive  employing a truncation mechanism while penalizing the prediction scores with a polynomial function. We highlight its significant advantages:

\begin{table}
\centering
\footnotesize
\setlength{\tabcolsep}{1.8mm}
\begin{tabular}{l|cc|cc|cc} 
\toprule
\multirow{2}{*}{Loss} & \multicolumn{2}{c|}{Gowalla} & \multicolumn{2}{c|}{AmazonKitchen} & \multicolumn{2}{c}{AmazonBeauty}  \\
                      & $k_1$    & $k_2$                         & $k_1$    & $k_2$                        & $k_1$     & $k_2$                  \\ 
\hline
SL                    & 187.85 & 110.95                     & 27.96 & 26.26                      & 24.84 & 22.48              \\
DrRL                  & 58.16  & 51.20                      & 20.60 & 18.68                      & 21.02 & 19.41              \\
\bottomrule
\end{tabular}
\caption{The column `$k_1$' denotes the average ratios of the largest weight over the average weights; `$k_2$' denotes ratios of the weights of false negative instances over the average weights.}
\label{table:noise to neg of SL}
\end{table}

\textbf{Subsumes SL and CCL.} $\lossName$ effectively combines the merits of both CCL and SL, incorporating the truncation mechanism from CCL and weighting strategies from SL. In fact, $\lossName$ can degenerate to SL and CCL under certain conditions, $\lossName$ $\to$ SL with $\wNormal \to 1$ and $\lossName$ $\to$ CCL with $\wNormal \to +\infty$ (see \Cref{proof of generalization of renyi}). This demonstrates that our $\lossName$ builds upon the foundations of SL and CCL, ensuring performance at least not worse than these existing methods.

\textbf{Robustness to False Negative Instances.} Unlike SL's exponential weighting, which can lead to disproportionate influence from false negatives due to its explosive nature, $\lossName$ offers a flexible polynomial weight distribution controlled by $\gamma$. The following lemma substantiates this:
\begin{lemma}
\label{lemma3: the worst-case distribution}
    $\lossName$ optimizes models under the following worst-case negative distribution:
    \begin{equation}
    \begin{aligned}
        \distFamily^*_{u}(j) =& \SLNegWeight\userNegDis(j), \\
        \SLNegWeight =& \negFac\frac{(\similarity(u,j) - \beta_u)_+^{\frac{1}{\wNormal - 1}}}{\mathbb{E}_{j \sim \userNegDis}\left[(\similarity(u,j) - \beta_u)_+^{\wStar}\right]^{\frac{1}{\wNormal}}} 
    \end{aligned}
    \end{equation}
    where  $\mathbb{E}_{j \sim \userNegDis}\left[\distFamily^*_{u}(j)\right] = 1$, $\SLNegWeight \propto (\similarity(u,j) - \beta_u)_+^{\frac{1}{\wNormal - 1}}$.
\end{lemma}
The proof is given in \Cref{appendix proof: the worst-case distribution}. The weight $\SLNegWeight$ is proportional to a polynomial function of prediction scores. The polynomial function is relatively milder and significantly mitigates the impact of false negatives, enhancing model robustness. Table~\ref{table:noise to neg of SL} provides the empirical evidence by comparing the weights of $\lossName$ with SL. We can find the weights of noisy data in $\lossName$ are significantly smaller than SL.

\textbf{Better Data Utilization.} Although $\lossName$ also employs a truncation mechanism, it boasts much better data utilization than CCL. Figure~\ref{fig:poor data utilization of CCL} compares the ratio of filtered instances between $\lossName$ and CCL across users. $\lossName$ filters significantly fewer instances --- on average, 10\% less than CCL.

This improvement can be attributed to the inclusion of the weighting strategy in $\lossName$. The OOD robustness of $\lossName$ can partly originate from the weighting strategy, reducing the need for a large truncation ratio to maintain OOD robustness. Figure~\ref{fig: user preference diff and cut trend} also provides empirical evidence that as $\gamma$ increases, the truncation ratio decreases.

\textbf{Learnable Personalized Margin Parameter $\beta$. } Eq.(\ref{eqs: close form renyi loss}) also provides a theoretical framework to learn the personalized margin parameter $\beta$. This approach avoids additional hyper-parameter tuning for $\beta$ and allows for the learning of fine-grained $\beta$, being potentially more effective given the diversity of user preference scores. Additionally, the objective with respect to $\beta$ is convex (\confer Appendix A.3), facilitating feasible and efficient learning. In practice, the model and $\beta$ can be updated iteratively and alternately via gradient descent, incurring minimal computational complexity.

\textbf{Easy Implementation and Minimal Hyper-parameter Tuning.} Appendix C details the implementation of $\lossName$. It can be straightforward to integrate into various recommendation models with minimal code modifications. Moreover, given $\beta$ is learnable, $\lossName$ requires minimal hyper-parameter tuning. In our experiments, in many cases, we find that simply setting $\negFac = 1$ yields satisfactory performance, leaving only $\wNormal$ as a parameter that needs tuning. Nevertheless, fine-tuning $\negFac$ would be better.

\section{Experiments}
\label{sec:experiment}

\begin{table*}[t]
\centering
\footnotesize
\setlength{\tabcolsep}{1.5mm}
\begin{tabular}{c|l|cc|cc|cc|cc} 
\toprule
\multirow{2}{*}{Backbone} & \multicolumn{1}{c|}{\multirow{2}{*}{Loss}} & \multicolumn{2}{c|}{Gowalla}                           & \multicolumn{2}{c|}{AmazonKitchen}                     & \multicolumn{2}{c|}{AmazonElectronics}                 & \multicolumn{2}{c}{AmazonBeauty}                       \\ 
\cline{3-10}
& \multicolumn{1}{c|}{}  & \multicolumn{1}{c}{Recall@20} & \multicolumn{1}{c|}{NDCG@20} & \multicolumn{1}{c}{Recall@20} & \multicolumn{1}{c|}{NDCG@20} & \multicolumn{1}{c}{Recall@20} & \multicolumn{1}{c|}{NDCG@20} & \multicolumn{1}{c}{Recall@20} & \multicolumn{1}{c}{NDCG@20}  \\ 
\hline
\multirow{10}{*}{MF}      
&MSE&$0.1326$&$0.1087$&$0.0259$&$0.0141$&$0.0744$&$0.0460$&$0.1276$&$0.0728$\\
&BCE&$0.1369$&$0.1065$&$0.0320$&$0.0162$&$0.0715$&$0.0444$&$0.1378$&$0.0781$\\
&BPR&$0.1441$&$0.1202$&$0.0260$&$0.0141$&$0.0663$&$0.0415$&$0.1150$&$0.0668$\\
&CCL&$\underline{0.1652}$&$\underline{0.1282}$&$\underline{0.0477}$&$0.0257$&$\underline{0.0870}$&$\underline{0.0554}$&$0.1518$&$0.0897$\\
&LLPAUC&$0.1444$&$0.1219$&$0.0362$&$0.0190$&$0.0793$&$0.0497$&$0.1506$&$0.0877$\\
&SL&$0.1623$&$0.1252$&$0.0453$&$\underline{0.0262}$&$0.0833$&$0.0513$&$0.1573$&$0.0941$\\
&BSL&$0.1613$&$0.1243$&$0.0454$&$0.0262$&$0.0835$&$0.0516$&$0.1587$&$\underline{0.0962}$\\
&AdvInfoNCE&$0.1628$&$0.1252$&$0.0443$&$0.0259$&$0.0815$&$0.0508$&$\underline{0.1596}$&$0.0952$\\
&\lossName&$\textbf{0.1785}$&$\textbf{0.1435}$&$\textbf{0.0501}$&$\textbf{0.0270}$&$\textbf{0.0895}$&$\textbf{0.0566}$&$\textbf{0.1645}$&$\textbf{0.0978}$\\

\cline{2-10}

&Imp.\%&$+8.06\%^*$&$+11.98\%^*$&$+5.04\%^*$&$+3.05\%^*$&$+2.92\%^*$&$+2.10\%^*$&$+3.13\%^*$&$+1.61\%^*$\\

\hline
\multirow{10}{*}{LightGCN}       
&MSE&$0.1471$&$0.1146$&$0.0329$&$0.0176$&$0.0781$&$0.0488$&$0.1375$&$0.0773$\\
&BCE&$0.1543$&$0.1303$&$0.0346$&$0.0189$&$0.0795$&$0.0504$&$0.1469$&$0.0836$\\
&BPR&$0.1550$&$0.1280$&$0.0359$&$0.0199$&$0.0797$&$0.0510$&$0.1435$&$0.0823$\\
&CCL&$0.1633$&$0.1281$&$\underline{0.0481}$&$\underline{0.0265}$&$\underline{0.0890}$&$\underline{0.0564}$&$0.1514$&$0.0881$\\
&LLPAUC&$\underline{0.1690}$&$\underline{0.1433}$&$0.0415$&$0.0228$&$0.0876$&$0.0554$&$0.1582$&$0.0926$\\
&SL&$0.1609$&$0.1245$&$0.0454$&$0.0258$&$0.0812$&$0.0502$&$\underline{0.1589}$&$\underline{0.0949}$\\
&BSL&$0.1611$&$0.1244$&$0.0457$&$0.0258$&$0.0822$&$0.0509$&$0.1556$&$0.0922$\\
&AdvInfoNCE&$0.1628$&$0.1251$&$0.0448$&$0.0254$&$0.0814$&$0.0504$&$0.1587$&$0.0944$\\
&\lossName&$\textbf{0.1788}$&$\textbf{0.1447}$&$\textbf{0.0502}$&$\textbf{0.0275}$&$\textbf{0.0902}$&$\textbf{0.0572}$&$\textbf{0.1665}$&$\textbf{0.0974}$\\

\cline{2-10}
&Imp.\%&$+5.79\%^*$&$+1.02\%^*$&$+4.40\%^*$&$+3.80\%^*$&$+1.36\%^*$&$+1.49\%^*$&$+4.76\%^*$&$+2.58\%^*$\\

\hline
\multirow{10}{*}{XSimGCL}      
&MSE&$0.1375$&$0.1093$&$0.0358$&$0.0199$&$0.0760$&$0.0494$&$0.1446$&$0.0827$\\
&BCE&$0.1530$&$0.1299$&$0.0356$&$0.0201$&$0.0782$&$0.0509$&$0.1496$&$0.0872$\\
&BPR&$\underline{0.1655}$&$\underline{0.1379}$&$0.0382$&$0.0210$&$0.0839$&$0.0539$&$0.1471$&$0.0859$\\
&CCL&$0.1614$&$0.1252$&$\underline{0.0472}$&$\underline{0.0258}$&$0.0876$&$\underline{0.0561}$&$0.1522$&$0.0891$\\
&LLPAUC&$0.1582$&$0.1340$&$0.0435$&$0.0252$&$\underline{0.0877}$&$0.0561$&$0.1490$&$0.0892$\\
&SL&$0.1508$&$0.1146$&$0.0413$&$0.0236$&$0.0748$&$0.0460$&$0.1511$&$0.0895$\\
&BSL&$0.1509$&$0.1144$&$0.0429$&$0.0243$&$0.0770$&$0.0465$&$0.1520$&$0.0898$\\
&AdvInfoNCE&$0.1531$&$0.1148$&$0.0427$&$0.0241$&$0.0771$&$0.0470$&$\underline{0.1526}$&$\underline{0.0907}$\\
&\lossName&$\textbf{0.1774}$&$\textbf{0.1417}$&$\textbf{0.0490}$&$\textbf{0.0266}$&$\textbf{0.0907}$&$\textbf{0.0578}$&$\textbf{0.1608}$&$\textbf{0.0953}$\\

\cline{2-10}
&Imp.\%&$+7.14\%^*$&$+2.81\%^*$&$+3.74\%^*$&$+3.04\%^*$&$+3.45\%^*$&$+2.96\%^*$&$+5.40\%^*$&$+5.14\%^*$\\

\bottomrule
\end{tabular}
\caption{Overall performance comparison of $\lossName$ with other losses. The best result is bolded and the runner-up is underlined. Imp.\% indicates the relatively improvements of $\lossName$ over the best baselines. The mark `*' suggests the improvement is statistically significant with $p<0.05$. }
\label{table: overall performance}
\end{table*}

\subsection{Experimental Setups}
\textbf{Datasets.} Following CCL and SL~\cite{mao2021simplex,wu2024bsl}, we adopt four widely-used datasets including AmazonKitchen, AmazonElectronics, AmazonBeauty~\cite{mcauley2015image} and Gowalla~\cite{he2020lightgcn}. The data information and preprocessing details are presented in Appendix B.

\textbf{Baseline Methods.} The following baselines are included: 1) \textbf{MSE}~\cite{pan2008one}, \textbf{BCE}~\cite{johnson2014logistic}, and \textbf{BPR}~\cite{rendle2009bpr}, three classical losses in RS; 2) \textbf{CCL} \cite{mao2021simplex}: a representative loss with leveraging truncation mechanism; 3) \textbf{SL} \cite{wu2024effectiveness}: a representative loss with softmax function; 4) \textbf{BSL} ~\cite{wu2024bsl}, \textbf{AdvInfoNCE} ~\cite{zhang2024empowering}: two SOTA losses, improving SL with leveraging DRO on positive side or integrating hardness-aware ranking mechanism; 5) \textbf{LLPAUC} ~\cite{shi2024lower}, the SOTA loss approximately optimizing the lower-left partial AUC. Following~\cite{zhang2024empowering}, all losses are integrated with three representative recommendation backbones: 1) the basic \textbf{MF}~\cite{rendle2009bpr}; 2) the graph-based model~\cite{kipf2016semi,wang2019neural,dong2021equivalence,wu2022graph,gao2022graph,chen2024macro}: \textbf{LightGCN}~\cite{he2020lightgcn} and the SOTA~\textbf{XSimGCL}~\cite{yu2023xsimgcl}.

\textbf{Hyper-parameter Settings.} To maintain fairness across all methods, we tune each method with a very fine granularity to ensure their optimal performance (see Appendix B).

\begin{table}[htbp]
\centering
\footnotesize 
\setlength{\tabcolsep}{0.8mm}
\begin{tabular}{l|cc|cc} 
\toprule
\multirow{2}{*}{Loss} & \multicolumn{2}{c|}{AmazonKitchen} & \multicolumn{2}{c}{AmazonBeauty}  \\ 
\cline{2-5}
                      & Recall@20 & NDCG@20                & Recall@20 & NDCG@20               \\ 
\hline
DrRL-w/o-LP           & 0.0482    & 0.0269                 & 0.1582    & 0.0960                \\
DrRL-w/o-P            & 0.0477    & 0.0265                 & 0.1579    & 0.0922                \\
DrRL                  & 0.0501    & 0.0270                 & 0.1645    & 0.0978                \\
\bottomrule
\end{tabular}
\caption{Ablation Study, we examine two variations of $\lossName$, where the personalized strategy ($\lossName$-w/o-P) and the learnable strategy ($\lossName$-w/o-LP) are omitted.}
\label{exp: ablation study}
\end{table}

\subsection{Experimental Results}

\textbf{Overall Performance Comparisons.} Table~\ref{table: overall performance} presents the performance of $\lossName$ compared with baselines. Overall, our $\lossName$ consistently outperforms all compared methods across all datasets and backbones. Especially in Gowalla, $\lossName$ achieves impressive improvements --- average 6.9\% and 5.2\% in terms of Recall@20 and NDCG@20 respectively on three backbones. These results demonstrate $\lossName$ can indeed mitigate weaknesses of SL and CCL, making our performance even surpassing other state-of-the-art loss functions by a significant margin. Besides, we find that DRO-enhanced losses (SL, BSL, CCL, AdvInfoNCE, $\lossName$) generally exhibit superior effectiveness. This outcome underscores the significance of OOD robustness in RS.

\begin{figure}[!t]
  \centering
  \includegraphics[width=1\linewidth]{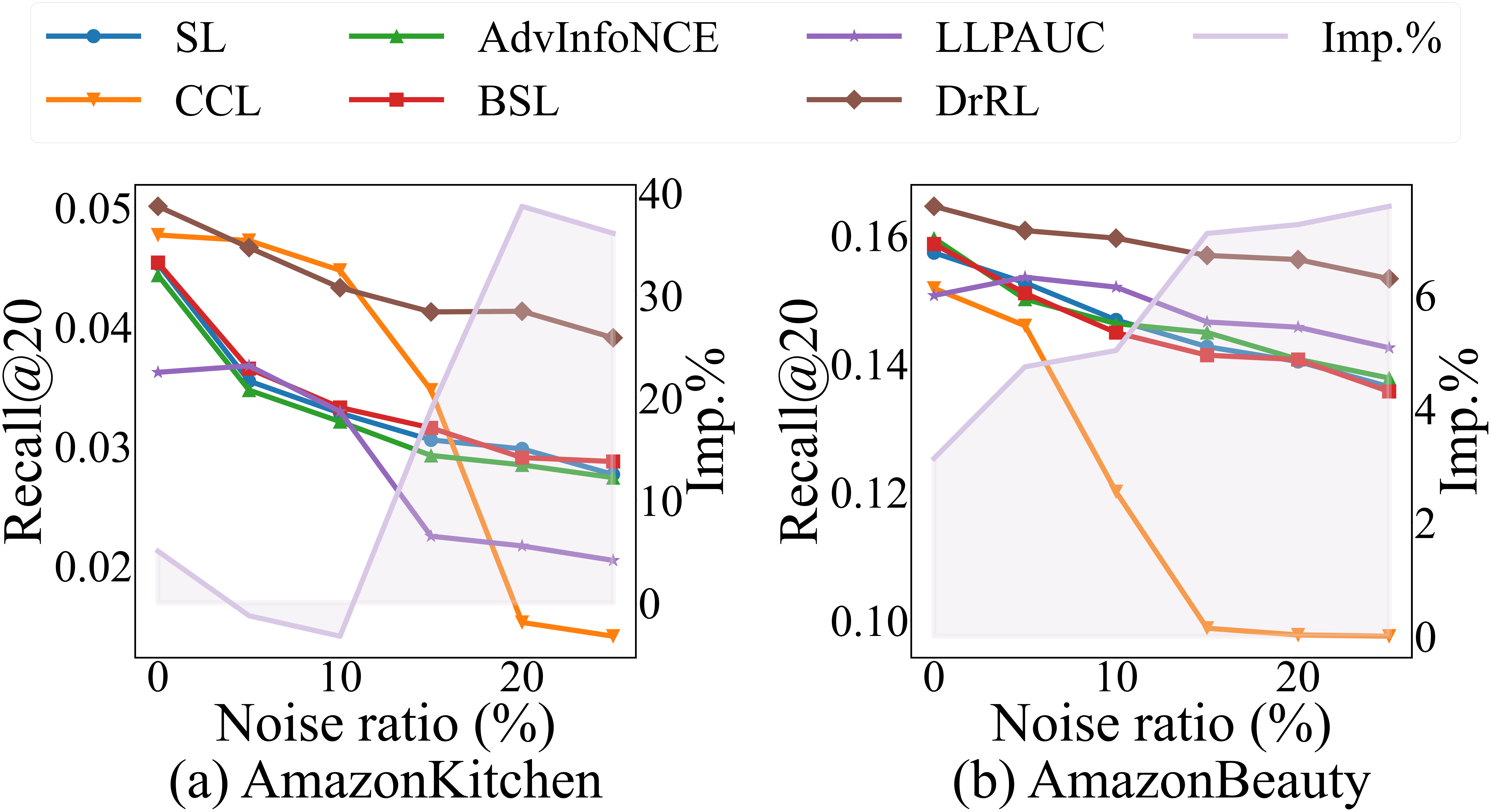}
  \caption{Performance comparisons with the varying ratios of false negative instances. We also present the relative improvements achieved by $\lossName$ over the best baselines.}
    \label{fig:noise-result-public}
\end{figure}

\textbf{Performance on Noisy Data.} Figure~\ref{fig:noise-result-public} illustrates the noise robustness of compared losses. Here, we closely follow \cite{wu2024bsl} and injects a certain proportion of positive items as false negative during negative sampling. As shown, $\lossName$ consistently outperforms the compared loss functions across most noise ratios. More notably, as the noise ratio increases, $\lossName$ demonstrates even greater improvements over the best baselines.

Another interesting observation is that while CCL exhibits a certain degree of noise robustness at lower noise ratios, its performance deteriorates rapidly as the noise level increases. This phenomenon can be explained as follows: CCL assigns uniform weights to the preserved negative instances rather than giving greater weight to those with higher prediction scores, which initially provides good robustness. However, as the noise ratio increases, the proportion of false negatives among the preserved instances significantly rises. Ultimately, this leads to a scenario where almost exclusively noisy data contributes to the training, resulting in the collapse of CCL’s performance under high noise conditions.

\begin{figure}[t!]
\centering
  \includegraphics[width=1\linewidth]{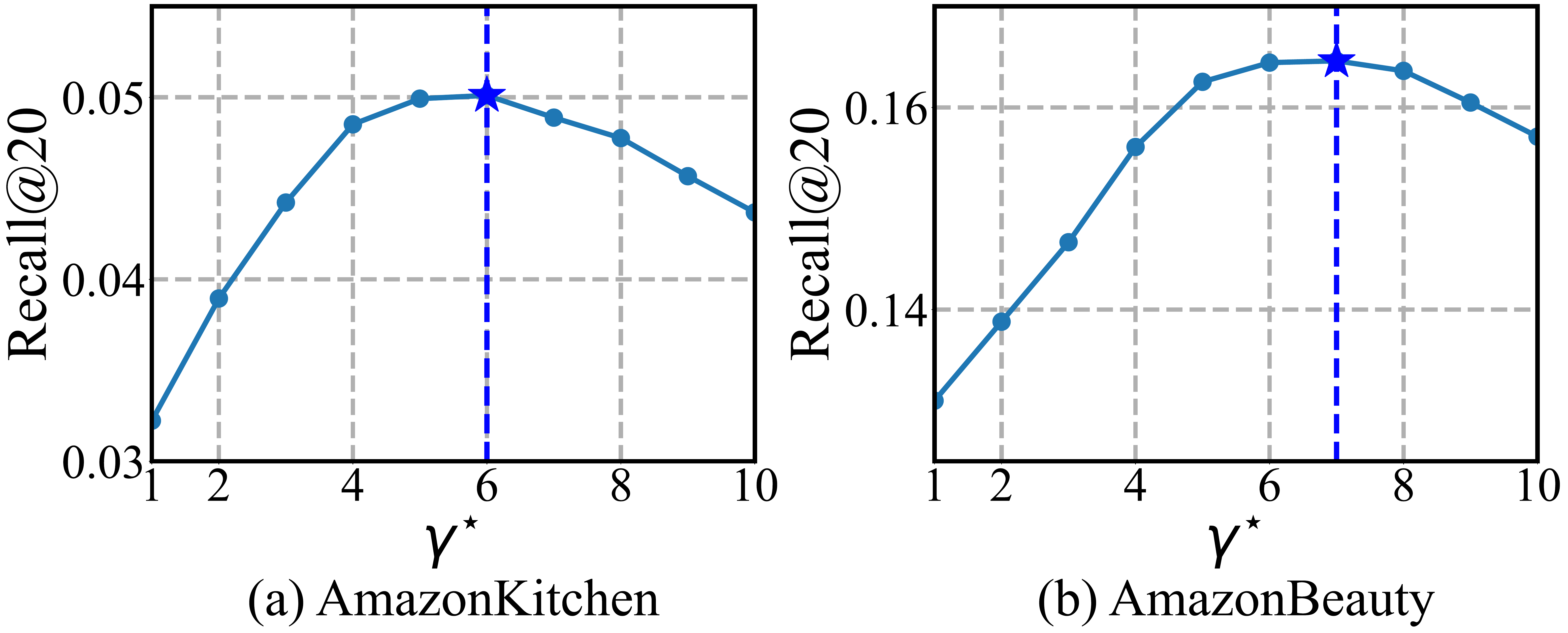}
  \caption{The impact of varying $\wNormal$.}
  \label{fig:case study}
\end{figure}

\textbf{Performance under Distribution Shifts.} To estimate the robustness of losses against distribution shifts, we follow \cite{wang2024distributionally} to construct OOD testing scenario --- we divide the training and testing dataset according to the interaction time, where the temporal bias is contained (\confer Appendix B). Table~\ref{exp: ood experiments} presents the results of various loss functions with temporal shifts. $\lossName$ achieves the best performance, demonstrating its superior OOD robustness. The effectiveness can be attributed to the incorporation of Rényi divergence. As a broader family of measures, including the divergences used in SL and CCL, Rényi divergence provides greater flexibility for adapting to complex OOD scenarios.

\textbf{Ablation Study.} In Table~\ref{exp: ablation study}, we evaluate two variants of $\lossName$, where the personalized strategy (P) and the learnable strategy (LP) are omitted (w/o). $\lossName$-w/o-LP outperforms $\lossName$-w/o-P by a large margin, demonstrating the necessity of leveraging personalized $\beta$. Additionally, we observe that $\lossName$-w/o-P performs slightly worse but remains close to $\lossName$-w/o-LP. This result indicates that our learning algorithm can indeed find an appropriate $\beta$, avoiding the need for exhaustive hyper-parameter tuning.

\textbf{The Impact of $\gamma$.} Figure~\ref{fig:case study} illustrates the model performance with $\wStar (\wStar = \frac{\wNormal}{\wNormal - 1})$. As shown, model performance initially increases with rising $\wStar$ but declines when $\wStar$ is further increased. This outcome corresponds with our theoretical expectations. Specifically, when $\wStar$ is set close to 1, $\lossName$ degenerates into the basic CCL; when $\wStar$ is set to a sufficiently large value, $\lossName$ approximates SL. Optimal performance is observed at an intermediate value of $\wStar$, where $\lossName$ effectively integrates the strengths of both SL and CCL, while substantially alleviating their respective limitations.  A larger $\wStar$ enhances data utilization as Figure~\ref{fig:case study} demonstrates, while smaller $\wStar$ flattens the weight distribution and enhances the noise robustness. Reflecting these dual effects, we observe a concave performance curve with respect to $\wStar$.

\begin{table}[t]
\centering
\footnotesize 
 \setlength{\tabcolsep}{0.8mm}
\begin{tabular}{l|cc|cc} 
\toprule
\multicolumn{1}{l|}{\multirow{2}{*}{Loss}} & \multicolumn{2}{c|}{AmazonKitchen} & \multicolumn{2}{c}{AmazonElectronics}   \\
\cline{2-5}
\multicolumn{1}{c|}{}                      & Recall@20 & NDCG@20                         & Recall@20 & NDCG@20                       \\ 
\hline
BPR&$0.01429$&$0.00749$&$0.04390$&$0.02718$\\
CCL&$ 0.02053 $&$ 0.01065 $&$ 0.05013 $&$0.03065$\\
LLPAUC&$0.01661$&$0.00860$&$0.04637$&$ 0.03135 $\\
SL&$0.01718$&$0.00914$&$0.04311$&$0.02508$\\
BSL&$0.01665$&$0.00907$&$0.04017$&$0.02377$\\
AdvInfoNCE&$0.01672$&$0.00892$&$0.04503$&$0.02554$\\
$\lossName$&$\textbf{0.02077}$&$\textbf{0.01114}$&$\textbf{0.05199}$&$\textbf{0.03160}$\\
\bottomrule
\end{tabular}
\caption{Performance comparisons under temporal shifts.}
\label{exp: ood experiments}
\end{table}

\section{Related Work}

As this work focuses on loss functions, here we mainly introduce related work on this topic (see Appendix D for recommendation models and recommendation robustness).

Loss functions have drawn increasing attention within the recommendation community. Initially, RS primarily utilized point-wise losses, treating recommendations as classification tasks, including MSE~\cite{pan2008one} and BCE~\cite{johnson2014logistic}. Subsequently, \citeauthor{rendle2009bpr} introduced pairwise loss through Bayesian personalized ranking~\cite{rendle2009bpr}. In more recent developments, there has been a surge in publications on loss functions. Notably, Softmax Loss (SL) \cite{wu2024effectiveness} and Cosine Contrastive Loss (CCL) \cite{mao2021simplex} are particularly effective. There are also some improvements over SL including BSL~\cite{wu2024bsl} that applies DRO on positive distributions, AdvInfoNCE~\cite{zhang2024empowering} that introduces a hardness-aware ranking mechanism, and BC \cite{zhang2023invariant} and PopDCL~\cite{liu2023popdcl} that integrate bias-aware terms.

Beyond these developments, explorations from other perspectives have also been conducted. For instance, automated machine learning has been employed to search for optimal loss functions among candidates \cite{li2022autolossgen}; Some researchers \cite{rashed2021guided,pulearning,shi2024lower} study surrogate objectives for NDCG or partial AUC metrics; while others \cite{wang2022towards,park2023toward} developed loss functions to enhance embedding alignment and uniformity.

This work focuses on analyzing CCL and SL, exploring their properties, and proposing an improved variant. The most closely related work is \cite{wu2024bsl}, but we have significant differences: 1) their analyses are merely on SL, while our analyses include both SL and CCL, elucidating their connections and limitations; 2) we introduce a novel loss that generalizes SL and CCL by incorporating Rényi divergence, distinctly different from their approach of integrating KL-divergence DRO in positive instances.

\section{Conclusions}

This work studies loss functions in recommender systems. Our comprehensive analyses have identified that both CCL and SL can be considered as enhancements of traditional losses through the application of DRO. However, their utilization of distinct distribution divergence metrics contributes to SL's high sensitivity to noise and CCL's low data utilization. To address these issues, this work introduces $\lossName$, which employs Rényi divergence within a DRO framework, effectively inheriting and enhancing the beneficial features of both SL and CCL while alleviating their drawbacks.

\section*{Acknowledgments}
This work is supported by the National Natural Science Foundation of China (62372399, 62476244), the Starry Night Science Fund of Zhejiang University Shanghai Institute for Advanced Study (SN-ZJU-SIAS001), OPPO Research Fund, and the advanced computing resources provided by the Supercomputing Center of Hangzhou City University.

\bibliography{aaai25}

\clearpage
\appendix
\newcommand{\ineqCons}{\lambda}
\newcommand{\eqCons}{\rho}

\section{Proof of Lemmas}

\subsection{Proof of Lemma~\ref{lem: CCL and SL DRO}}\label{appendix-CCL-and-SL}
In this section, we show that optimizing CCL and SL is equivalent to optimizing Eq.(\ref{eqs:Distributionally Contrastive Loss}) with different divergence constraints. Given that $f(u,i)$ serves as an constant do not affect the optimization objective, the KL-divergence constrained optimization problem can be reformulated as:
\begin{equation}\label{eqs:KL-constrained-DRO}
\begin{aligned}
    &\max\limits_{\userUncertainDis} \E_{j \sim \userUncertainDis}\left[f(u,j)\right] \\
    &D_{\text{KL}}(\userUncertainDis,\userNegDis) := \sum\limits_{j \in \negU} \userUncertainDis(j) \log\frac{\userUncertainDis(j)}{\userNegDis(j)} \le \eta.
\end{aligned}
\end{equation}
The equivalence between the optimization problem in Eq.(\ref{eqs:KL-constrained-DRO}) and SL is proved by \citeauthor{wu2024understanding} (see Appendix A.1 in \cite{wu2024understanding} for more details). Similarly, the worst-case regret divergence constrained optimization problem can be reformulated as:
\begin{equation}\label{eqs:worst-case-constrained-DRO}
\begin{aligned}
    &\max\limits_{\userUncertainDis} \E_{j \sim \userUncertainDis}\left[f(u,j)\right] \\
    & D_{\text{WR}}(\userUncertainDis, \userNegDis) := \esssup\limits_{j \in \negU}\log\left( \frac{\userUncertainDis(j)}{\userNegDis(j)}\right) \le \log(\cclWeight),
\end{aligned}
\end{equation}
The equivalence between the optimization problem in Eq.(\ref{eqs:worst-case-constrained-DRO}) and CCL is given by~\citeauthor{duchi2021learning} (see Example 3 in \cite{duchi2021learning} for more details). We also refer readers to Eq.(\ref{eqs: optimal beta of renyi loss}), CCL is a special case when $\gamma = +\infty$, which means the WR divergence constraint.

\subsection{Proof of Lemma~\ref{lemma2: renyi loss}}
\label{proof: Appendix 2}
In this section, we show the equivalence between the optimization problem in Eq.(\ref{eqs:renyi optimization objective}) and the closed form in Eq.(\ref{eqs: close form renyi loss}). Given that $\similarity(u,i)$ as constants do not affect the optimization, the aforementioned optimization objective in Eq.(\ref{eqs:renyi optimization objective}) can be simplified to the following form:
\begin{equation}
\begin{aligned}
    &\max\limits_{\userUncertainDis} \E_{j \sim \userUncertainDis}\left[f(u,j)\right] \\
    &\st  D_{\wNormal}(\userUncertainDis,\userNegDis) := \sum\limits_{j \in \negU} \userNegDis(j)\phi_{\wNormal}\left(\frac{\userUncertainDis(j)}{\userNegDis(j)}\right)  \le \eta \\
    & \phi_{\wNormal}(t) = \frac{1}{\wNormal(\wNormal - 1)}(t^{\wNormal} - \wNormal t + \wNormal - 1)\end{aligned}
\end{equation}
We focus on how to eliminate the inner maximization optimization problem and the distributional constraint term with Rényi divergence. Suppose $L(j) = \userUncertainDis(i) / \userNegDis(j)$, then the Rényi Divergence can be written as $\E_{\userNegDis}[\phi_{\wNormal}(L)]$. As a result, the inner maximization problem can be reformulated as follows:
\begin{equation}
    \begin{aligned}
            &\max\limits_{L} \E_{j \sim P_u^-}\left[f(u,j)L\right] \\
            &\st \E_{\userNegDis}[\phi_{\wNormal}(L)] \le \eta, \quad \E_{\userNegDis}\left[L\right] = 1
    \end{aligned}
\end{equation}
As a convex optimization problem, we use \textit{Lagrangian function} to solve it:
\begin{equation}
\label{eqs: level one of renyi loss proof}
\resizebox{1\hsize}{!}{$
\begin{aligned}
    &\min\limits_{\ineqCons \ge 0,\eqCons}\max\limits_{L} \left\{\E_{\userNegDis}\left[\similarity(u,j)L\right] -\ineqCons(\E_{\userNegDis}[\phi_{\wNormal}(L)] - \eta) - \eqCons(\E_{\userNegDis}[L] - 1)\right\} \\
    &= \min\limits_{\ineqCons \ge 0,\eqCons} \left\{ \ineqCons\eta + \eqCons +\ineqCons \max\limits_{L} \E_{j \sim \userNegDis}\left\{\frac{\similarity(u,j) - \eqCons}{\ineqCons}L - \phi_{\wNormal}(L)\right\} \right\} \\
    &= \min\limits_{\ineqCons \ge 0,\eqCons} \left\{ \ineqCons\eta + \eqCons +\ineqCons  \E_{j \sim \userNegDis}\left[
    \max\limits_{L}\left(\frac{\similarity(u,j) - \eqCons}{\ineqCons}L - \phi_{\wNormal}(L)\right)
    \right]\right\}
\end{aligned}
$}
\end{equation}
Notice that $\max_{L}\left(\frac{\similarity(u,j) - \eqCons}{\ineqCons}L - \phi_{\wNormal}(L)\right) = \phi_{\wNormal}^*\left(\frac{\similarity(u,j) - \eqCons}{\ineqCons}\right)$
is the \textit{Fenchel Conjugate Function} of $\phi_{\wNormal}(t)$, we have $\phi_{\wNormal}^*(x) = \frac{1}{\wNormal}\left((\wNormal - 1)x + 1\right)_+^{\wStar} - \frac{1}{\wNormal}$ (\confer Duchi \excetra~\cite{duchi2021learning}). Then, Eq.(\ref{eqs: level one of renyi loss proof}) can be derived as:
\begin{equation}
\label{eqs: level two renyi loss proof}
\begin{aligned}
    &\min\limits_{\ineqCons \ge 0,\eqCons} \left\{ \ineqCons\eta + \eqCons +\ineqCons  \E_{j \sim \userNegDis}\left[
    \max\limits_{L}\left(\frac{\similarity(u,j) - \eqCons}{\ineqCons}L - \phi_{\wNormal}(L)\right)\right]\right\} \\
    &= \min\limits_{\ineqCons \ge 0,\eqCons} \left\{ \ineqCons\eta + \eqCons +\ineqCons  \E_{j \sim \userNegDis}\left[
    \phi^*_{\wNormal}\left(\frac{\similarity(u,j) - \eqCons}{\ineqCons}\right)\right]\right\} \\
\end{aligned}   
\end{equation}
According to Duchi\excetra~\cite{duchi2021learning}, with $\beta = \eqCons - \frac{\ineqCons}{\wNormal - 1}$, the optimal $\ineqCons^*$ which minimize the preceding expression is:
\begin{equation}
\label{eqs: the optimal inequality coef}
    \ineqCons^* = (\wNormal - 1)(\wNormal(\wNormal - 1)\eta + 1)^{-\frac{1}{\wStar}}\E_{j \sim \userNegDis}\left[(\similarity(u,j) - \beta)_+^{\wStar}\right]^{\frac{1}{\wStar}},
\end{equation}
where $\wStar = \frac{\wNormal}{\wNormal - 1}$. Ultimately, substituting $\phi^*_{\wNormal}(x)$, $\ineqCons^*$ and $\beta = \eqCons - \frac{\ineqCons}{\wNormal - 1}$ back into Eq.(\ref{eqs: level two renyi loss proof}), we have: 
\begin{equation}
\label{eqs: optimal beta of renyi loss}
    \min\limits_{\beta} \left\{ \negFac \E_{j \sim \userNegDis}\left[(\similarity(u,j) - \beta)_+^{\wStar} \right]^{\frac{1}{\wStar}} + \beta \right\},
\end{equation}
where $\negFac = (\wNormal(\wNormal - 1)\eta + 1)^{\frac{1}{\wNormal}}$. Thus Lemma~\ref{lemma2: renyi loss} is proven. 

\subsection{Proof of Lemma~\ref{lemma3: the worst-case distribution}}
\label{appendix proof: the worst-case distribution}
For the inner optimization in Eq.(\ref{eqs: level one of renyi loss proof}), we can draw the optimal $L^*$ for $\max_{L}\left(\frac{\similarity(u,j) - \eqCons}{\ineqCons}L - \phi_{\wNormal}(L)\right)$ as:
\begin{equation}
    \frac{\similarity(u,j) - \eqCons^*}{\ineqCons^*} = \phi_{\wNormal}'(L^*),
\end{equation}
where $\ineqCons^*$ and $\eqCons^*$ denote the optimal coefficients that minimize Eq.(\ref{eqs: level one of renyi loss proof}). With $\phi_{\wNormal}'(L) = \frac{1}{\wNormal - 1}(L^{\wNormal - 1} - 1)$ and $\eqCons^* = \beta^* + \frac{\ineqCons^*}{\wNormal - 1}$, we have:
\begin{equation}
    \similarity(u,j) - \beta^* = \frac{\ineqCons^*}{\wNormal - 1}(L^*)^{\wNormal - 1}
\end{equation}
Substituting $\ineqCons^*$ in Eq.(\ref{eqs: the optimal inequality coef}), we have:
\begin{equation}
\begin{aligned}
    L^* =& \negFac\frac{(\similarity(u,j) -  \beta^*)_+^{\frac{1}{\wNormal - 1}}}{   E_{j \sim \userNegDis}[(\similarity(u,j) -  \beta^*)_+^{\wStar}]^{\frac{1}{\wNormal}} } \\
    Q_u^*(j) =&  \negFac\frac{(\similarity(u,j) - \beta^*)_+^{\frac{1}{\wNormal - 1}}}{   E_{j \sim \userNegDis}[(\similarity(u,j) - \beta^*)_+^{\wStar}]^{\frac{1}{\wNormal}} } \userNegDis(j)
\end{aligned}
\end{equation}
Taking derivative with respect to $\beta$ to minimize Eq.(\ref{eqs: optimal beta of renyi loss}), we have:
\begin{equation}
    \E_{j \sim \userNegDis}\left[\negFac\frac{(\similarity(u,j) - \beta^*)_+^{\frac{1}{\wNormal - 1}}}{   E_{j \sim \userNegDis}[(\similarity(u,j) - \beta^*)_+^{\wStar}]^{\frac{1}{\wNormal}} }\right] = 1
\end{equation}
Thus, we can get $\E_{j \sim \userNegDis}\left[ Q_u^{*}(j)\right] = 1$. The Lemma~\ref{lemma3: the worst-case distribution} is proven.

\subsection{The Convexity with Respect to $\beta$}
\label{section proof: convexity of beta}
Let $h(\beta) = \negFac \E_{j \sim \userNegDis}\left[(\similarity(u,j) - \beta)_+^{\wStar} \right]^{\frac{1}{\wStar}} + \beta$, we have:
\begin{equation}
\resizebox{1\hsize}{!}{$
\begin{aligned}
    &h(k \beta_1 + (1-k)\beta_2) =k\beta_1 + (1-k)\beta_2 \\
    &+\negFac  \E_{j \sim \userNegDis}\left[\bigg(k(\similarity(u,j) - \beta_1) + (1- k )(\similarity(u,j) - \beta_2)\bigg)_+^{\wStar} \right]^{\frac{1}{\wStar}}
\end{aligned}
$}
\end{equation}
With \textit{Minkowski’s inequality}~\cite{minkowski1910geometrie}, we have:
\begin{equation}
\begin{aligned}
    &\E_{j \sim \userNegDis}\left[\bigg(k(\similarity(u,j) - \beta_1) + (1- k )(\similarity(u,j) - \beta_2)\bigg)_+^{\wStar} \right]^{\frac{1}{\wStar}} \\
    &\le k\E_{j \sim \userNegDis}\left[(\similarity(u,j) - \beta_1)_+^{\wStar} \right]^{\frac{1}{\wStar}} \\
    &+ (1- k)\E_{j \sim \userNegDis}\left[(\similarity(u,j) - \beta_2)_+^{\wStar} \right]^{\frac{1}{\wStar}}
\end{aligned}
\end{equation}
Thus, we have:
\begin{equation}
    h(k \beta_1 + (1-k)\beta_2) \le k h( \beta_1) + (1-k)h(\beta_2)
\end{equation}
The convexity with respect to $\beta$ is proven.

\subsection{Generalizing SL and CCL}
\label{proof of generalization of renyi}
It is evident that $\lossName$ is the same as CCL in terms of $\wStar = 1 \ (\wNormal \to +\infty)$. Furthermore, KL divergence is a special case of Rényi divergence in terms of $\wNormal = 1$. Following Wu \excetra~\cite{wu2024understanding}, $\lossName$ also generalizes SL.

\section{Experimental Details}
\label{appendix dataset}
\textbf{Datasets.} The statistics of each dataset used in the experiments are shown in Table~\ref{tab:dataset statistics}, which lists the number of users, items interactions and sparsity of the dataset. The Gowalla and AmazonBeauty is extracted from the open source repository~\footnote{\url{https://huggingface.co/reczoo} }. We further pre-process the AmazonKitchen and AmazonElectronics~\footnote{\url{https://cseweb.ucsd.edu/~jmcauley/datasets/amazon/links.html}}. Following Wang \excetra~\cite{wang2019neural} and He \excetra~\cite{he2016vbpr}, we use 10-core setting and 5-core setting for AmazonElectronics and AmazonKitchen, $\ie$all users and items have at least 10-interactions and 5-interactions, respectively. Following He~\excetra~\cite{he2020lightgcn}, we adopt the early stopping strategy, in which premature stop if NDCG@20 on the validation set does not increase for 25 successive epochs. The detailed dataset constructions in IID, OOD, and Noise setting are as follows:
\begin{itemize}
    \item \textbf{IID:} In the IID setting, we randomly partition the original dataset into training and test sets that maintain identical distributions.  Specifically, the user-item interactions are i.i.d split into 80\% training and 20\% test set. Furthermore, 10\% of training set are split as validation set. In the IID setting, the training and test sets are both long-tail.

    \item \textbf{OOD:} In the OOD setting, we follow~\cite{wang2024distributionally}, processing AmazonKitchen and AmazonElectronics to exhibit temporal shift (As there exists no timestamp in Gowalla and AmazonBeauty in \url{https://huggingface.co/reczoo}, we only utilize AmazonKitchen and AmazonElectronics to conduct temporal shift). Specifically, we take the most recent 20\% of interaction data from each user as the test set, and the earliest 80\% of interactions data as the training set. Similarly, 10\% of training set are split as validation set. We further filter the items that not exist in training set, in which the item number would decrease slightly in Table~\ref{tab:dataset statistics}. 
    
    \item \textbf{Noise:} The Noise setting retains the same validation/test splits as the IID setup. However, we introduce false negatives as noise, \ie a portion of training positive items are treated as negative. Specifically, the negative items will be sampled from the false negatives with a probability of $p$ as the negative noise~\cite{wu2024bsl}, where $p \in \{0.05,0.1,0.15,0.2,0.25\}$ is the noise ratio. 
\end{itemize}

\begin{table}[ht]
\caption{Dataset statistics.}
\label{tab:dataset statistics}
\resizebox{0.48\textwidth}{!}{
\begin{tabular}{@{}c|c|cccc@{}}
\toprule
Setting                          & Dataset      & \#Users & \#Items & \#Interactions & Density \\ \midrule
\multirow{4}{*}{IID Setting}
& Gowalla      & 29858   & 40981   & 1027370        & 0.0007                       \\
& AmazonKitchen       & 51552   & 23158    & 427744       & 0.0003                \\
& AmazonElectronics  & 13455   & 8360   & 234521        & 0.0018                 \\
& AmazonBeauty     & 7068   & 3570   & 79506        & 0.0027                   \\ \midrule
\multirow{2}{*}{Temproal Shift}
& AmazonKitchen             & 51552    & 23136  & 402119         & 0.0002          \\
& AmazonElectronics         & 13455    & 8323   & 223595         & 0.0017           \\ \midrule
\end{tabular}
}
\end{table}

\subsubsection{Metrics.} We follow~\cite{wu2024bsl,zhang2024empowering}, adopting two representative evaluation metrics, NDCG@$20$ and Recall@$20$ for performance evaluation.

\subsubsection{Baseline Methods.} To demonstrate the superiority of our proposed method, we compare the classical and state-of-the-art recommendation loss functions, including:

\begin{itemize}
    \item \textbf{BCE} (NeurIPS'14~\cite{johnson2014logistic}), \textbf{MSE} (ICDM'08~\cite{pan2008one}) and \textbf{BPR} (UAI' 09~\cite{rendle2009bpr}): three classical and conventional loss functions in RS, which treat recommendations as classification or Beyasian personalized ranking.
    \item \textbf{CCL} (CIKM'21~\cite{mao2021simplex}): a representative loss function with leveraging truncation mechanism. 
    \item \textbf{SL} (TOIS'24~\cite{wu2024effectiveness}): a representative loss function that normalizes model predictions with softmax function and enhances the score of positive as compared with negatives. 
    \item \textbf{BSL} (ICDE'24~\cite{wu2024bsl}), \textbf{AdvInfoNCE} (NeurIPS' 24~\cite{zhang2024empowering}): two state-of-the-art loss functions on the basis of SL, which further leverages DRO on positive instances, or integrates fine-grained hardness-aware ranking strategy, respectively. 
    \item \textbf{LLPAUC} (WWW'24~\cite{shi2024lower}), the state-of-the-art loss function that approximately optimizes the lower-left partial AUC.
\end{itemize}

\subsection{Backbone Settings}

We implemented three popular recommendation backbones in our experiments, including:
\begin{itemize}
    \item \textbf{MF}~\cite{rendle2009bpr}: MF serves as a foundational framework in recommendation systems, which decomposes the user-item interaction matrix into low-dimensional user and item embeddings. Modern embedding-based recommenders uniformly adopt MF as the first layer. In this work, we follow Wang \excetra~\cite{wang2019neural}, setting the embedding size $d = 64$ for all recommendation backbones.
    
    \item \textbf{LightGCN}~\cite{he2020lightgcn}: LightGCN is a simplified GNN-based recommendation model, which removes the nonlinear activations and feature transformations in NGCF~\cite{wang2019neural}. By performing non-parameterized graph convolution, LightGCN achieves superior recommendation performance. In this work, we follow the original setting in He \excetra~\cite{he2020lightgcn}, setting the graph convolution number as 2.
    
    \item \textbf{XSimGCL}~\cite{yu2023xsimgcl}: XSimGCL advances recommendation systems by integrating contrastive learning into graph aggregation for user-item interactions. Specifically, XSimGCL perturbs the output embeddings at each layer with stochastic noise, and introduces contrastive learning by incorporating an auxiliary InfoNCE~\cite{oord2018representation} loss between the final layer and the $l^*$ layer. In this work, we follow Yu \excetra~\cite{yu2023xsimgcl}, choosing  the 2-layer LightGCN-based model. Moreover, the contrastive layer $l^* = 1$ (where the embedding layer is 0-th layer), the temperature of InfoNCE is set as $0.2$, and the modulus of random noise to $0.2$. Furthermore, we tune the model, setting the weight of auxiliary InfoNCE as $0.001$, which ensures the performance of all methods.

\end{itemize}

\subsection{Hyper-parameter Settings}

\textbf{Optimizer.} We use Adam~\cite{kingma2014adam} optimizer for training (LLPAUC is followed by~\cite{shi2024lower} with its defined min-max Adam). The learning rate (lr) is searched in $\{1e^{-1},1e^{-2},1e^{-3},1e^{-4},1e^{-5}\}$, except for MSE, BCE and MSE, where the lr is searched in $\{1e^{-1},1e^{-2},...,1e^{-8}\}$. The weight decay (conducted on normalized user and item embeddings if similarity score is adopted) is searched in $\{1e^{-1},1e^{-2},...,1e^{-8}\}$. The batch size is set as 1024, and we uniformly sample 1024 negative items for each positive interaction in training. 

The hyper-parameter of each loss are detailed as follows:

\noindent\textbf{BPR, MSE, and BCE:} No other hyper-parameters.

\noindent\textbf{LLPAUC:} We keep the original setting in Shi~\cite{shi2024lower}, and further highly fine-tune the hyper-parameters $\alpha \in \{0.01,0.05,0.10,...,1.6\}$ and $\beta \in \{0.01,0.02,...,1.0\}$.

\noindent\textbf{Softmax:} The temperature $\tau \in \{0.05,0.06,...,0.40\}$.

\noindent\textbf{CCL:} Following Mao \excetra~\cite{mao2021simplex}, the weight factor $\alpha \in \{1,2,...,15\}$. In Gowalla, we tune $\alpha \in \{100,105,...,150\}$. The margin is searched in $\beta \in \{0.10,0.15,...,0.90\}$.

\noindent\textbf{AdvInfoNCE:} The temperature is searched in the same space as SL. Following Zhang \excetra~\cite{zhang2024empowering}, we set the negative weight as 64, the adversarial learning rate is searched in $\{1e^{-1},1e^{-2},...,1e^{-6}\}$. The adversarial training epochs $E_{adv}$ are searched in $\{1,2,..,30\}$, and the adversarial training interval $T_{adv}$ is searched in $\{5,10,...,30\}$.

\noindent\textbf{BSL:} Following Wu \excetra~\cite{wu2024bsl}, we tune $\tau_2$ as searched in the same space as SL, and $\frac{\tau_2}{\tau_1} \in \{0.60,0.65,...,1.60\}$. 

\noindent\textbf{$\lossName$:} We tune the initial value $\beta_0$ as the same space with CCL. The margin learning rate is searched in $\{1e^{-4},1e^{-5}\}$. The margin learning is optimized with SGD optimizer. In IID setting, we set $\negFac = 1$, and tune $\wNormal \in \{1.05,1.06,...,1.35\}$. In OOD setting, we tune $\negFac \in \{1.05,1.10,...,1.50\}$ and $\wNormal \in \{1.05,1.06,...,1.35\}$, and plus the setting $\negFac \in \{1,2,...,15\}$, $\wNormal \in \{1.5,2.0,...,7.0\}$.
\section{Optimization strategies}
\label{appendix: the learning algorithm of drrl}

\subsection{The Algorithm of Learnable Personalized Margin Parameter}
Setting user personalized margin parameters $\boldsymbol{\beta} = [\beta_1,\beta_2,...\beta_{\vert \userSet \vert}]^{\top}$, and:
\begin{equation}
    \loss_{\beta}(u) = \negFac  \left[\frac{1}{\vert \negU \vert} \sum\limits_{j \in \negU}(\similarity(u,j) - \normalBeta_u)_+^{\wStar}\right]^{\frac{1}{\wStar}} + \normalBeta_u.
\end{equation}
The optimization algorithm for $\lossName$ is as follows:

\begin{algorithm}[H]
    \caption{Learning the personalized margin of $\lossName$}
    \begin{algorithmic}[1]
    \label{algorithm}
    \STATE \textbf{Input}: {User set $\mathcal{U}$, Item set $\mathcal{I}$, learning parameters $\{\theta, \beta_u\}$}, model learning rate $lr$, margin learning rate $lr_{\beta}$, parameter $\wNormal$.
    \STATE \textbf{Initialize}: Initialize the $\beta_u$ as $\beta_0$ for each user. 
   \REPEAT
        \STATE Sample a mini-batch positive interaction $\mathcal{B}^+$
        
        \STATE Uniformly sample a mini-batch $\mathcal{B}_u^-\in\mathcal{I}_u^-$ for each $(u,i)\in \mathcal{B}^+$.

        \STATE Compute $\loss_{\boldsymbol{\beta}} = \sum_{(u,i) \in \mathcal{B}^+} \loss_{\beta}(u)$;
        \STATE Update $\boldsymbol{\beta} \leftarrow lr_{\beta} \cdot (\partial \loss_{\boldsymbol{\beta}}  / \partial \boldsymbol{\beta}) $;

        \STATE Compute $\loss_{\theta} = \vert \mathcal{B}^+\vert^{-1}\sum_{(u,i) \in \mathcal{B}^+} \loss_{\lossName}(u)$ in Eq.(\ref{eqs: close form renyi loss})
        
        \STATE Update $\theta \leftarrow lr_{\theta}\cdot(\partial \loss_{\theta} / \partial \theta)$;
   \UNTIL{CF model converges}
    \end{algorithmic}
\end{algorithm}

\subsection{Implementation Details}
We add a small $\text{eps} = 1e^{-1}$ to avoid the numerical overflow when $\wNormal$ is small, and adopt the structure:
\begin{equation}
\begin{aligned}
    &\loss_{\lossName}(u) = -\frac{1}{\vert \posU \vert}\sum\limits_{i \in \posU}\similarity(u,i) + \\ 
    &\left(\frac{1}{\vert \negU \vert}\sum\limits_{j \in \negU}\left[\negFac (\similarity(u,j) - \beta)_+ + \text{eps}\right]^{\wStar}\right)^{\frac{1}{\wStar}},
\end{aligned}
\end{equation}
and 
\begin{equation}
\begin{aligned}
    &\loss_{\beta}(u) = \normalBeta_u + \\
    &\left(\frac{1}{\vert \negU \vert}\sum\limits_{j \in \negU}\left[\negFac (\similarity(u,j) - \beta)_+ + \text{eps}\right]^{\wStar}\right)^{\frac{1}{\wStar}},
\end{aligned}
\end{equation}
in the practical training. All experiments are conducted on one RTX 2080TI GPU and one Intel(R) Xeon(R) Gold 6254 @ 3.10GHz 18--Core Processor.
\section{Detailed Related Work}
\label{apdix: detailed related work}

\subsection{Recommendation Models}
Recently, researchers have shown a surge of interests on designing recommendation models for collaborative filtering. Initial studies centered on improving \textit{Matrix Factorization} to model more sophisticated user-item interactions. The representative model architectures in this type includes MF~\cite{koren2009matrix}, LRML~\cite{tay2018latent}, SVD~\cite{deerwester1990indexing,bell2007modeling}, SVD++~\cite{koren2008factorization}, NCF~\cite{he2017neural}, $\etc$. Inspired by the success of Graph Neural Networks (GNNs,~\cite{wu2022graph,gao2022graph,kipf2016semi,wang2019neural}) in capturing complex connections between users and items, GNN-based recommenders including LightGCN~\cite{he2020lightgcn}, NGCF~\cite{wang2019neural}, LCF~\cite{yu2020graph}, and APDA~\cite{zhou2023adaptive} have gained prominence. Recent studies have explored integrating the contrastive learning paradigm  (\eg SGL~\cite{wu2021self}, XSimGCL~\cite{yu2023xsimgcl}) with LightGCN, which not only enhances graph representations, but also achieves SOTA recommendation performance.

\subsection{Recommendation Robustness}

This work mainly focus on enhance model robustness under noise and OOD scenario, thus we review studies in these two topics. In OOD scenario, \cite{o2004efficient} initially incorporated neighbor selection into MF. Building upon causal learning, COR~\cite{wang2022causal} formalizes interactions via causal graph structures and leverages counterfactual reasoning to mitigate effects arising from OOD interactions. InvPref~\cite{wang2023efficient} and HIRL~\cite{zhang2023hierarchical} assign environment-specific variables to individual user-item interactions, splitting the dataset into diverse environments, and apply invariant learning to detect elements that persist consistently across varying environments. CausPref~\cite{he2022causpref} leverages a differentiable causal graph learning, which aims at capturing the invariant user preferences.

In noise scenario, some works~\cite{feng2021robust,yuan2021dual} leverage attention mechanism, while \citeauthor{chen2021structured}~\cite{chen2021structured} conduct denoising with stochastic binary masks. Furthermore, BERD~\cite{sun2021does} captures the interaction-specific uncertainty, and discards high-loss, low-uncertainty components for denoising. AutoDenoise~\cite{lin2023autodenoise} dynamically predict and select noise-free data. HSD~\cite{zhang2022hierarchical} measures inter-item similarities and items' alignment with user interests to perform noise detection. STEAM~\cite{lin2023self} employs a dual-model strategy, training a discriminator model and generator model, thereby safeguarding noise.

\section{Further Ablation Studies}
\subsection{Further Abldation Studies}
The ablation studies conducted on AmazonElectronics and Gowalla are shown in Table~\ref{tab: more experiments}. Similar with the ablation studies in section 5, it also demonstrates that $\lossName$-w/o-LP outperforms $\lossName$-w/o-P by a large margin, demonstrating the necessity of leveraging personalized $\beta$. Additionally, $\lossName$-w/o-P performs slightly worse but remains close to $\lossName$-w/o-LP, indicating that our learning algorithm can indeed find an appropriate $\beta$, avoiding the need for exhaustive hyper-parameter tuning.

\begin{table}[ht]
\caption{Ablation Study in AmamzonElectronics and Gowalla.}
\centering
\resizebox{0.48\textwidth}{!}{
\begin{tabular}{l|cc|cc} 
\toprule
\multirow{2}{*}{Loss} & \multicolumn{2}{c|}{Gowalla} & \multicolumn{2}{c}{AmazonElectronics}  \\ 
\cline{2-5}
                      & Recall@20 & NDCG@20                & Recall@20 & NDCG@20               \\ 
\hline
DrRL-w/o-LP           & 0.1717    & 0.1357                 & 0.0889    & 0.0560                \\
DrRL-w/o-P            & 0.1692    & 0.1325                  & 0.0895    & 0.0565                \\
DrRL                  & 0.1785    & 0.1435                 & 0.0895     & 0.0566                \\
\bottomrule
\end{tabular}}
\label{tab: more experiments}
\end{table}

\section{Optimal Hyper-parameters}

We report the optimal hyperparameters of each method on each dataset and backbone from \Cref{tables:OOD-hyperparameter,table:IID-Hyperparameter-official}, in the order of the hyperparameters listed in \Cref{tab:hyperparameters-searching}.

\begin{table}[H]
    \centering
    \caption{Hyperparameters to be searched for each method.}
    \begin{tabular}{l|l}
        \Xhline{1.3pt}
        \multicolumn{1}{c|}{\textbf{Method}} & \multicolumn{1}{c}{\textbf{Other Hyperparameters}} \bigstrut\\
        \Xhline{1pt}
        MSE         & no other hyperparameters                                                  \bigstrut[t]\\
        BCE        &   no other hyperparameters                                        \\
        BPR         &   no other hyperparameters                       \\
        CCL     &  \{$\alpha$, $\beta$\}                                                     \bigstrut\\
        LLPAUC  &  \{$\alpha$, $\beta$\}                                                               \bigstrut\\
        SL      & \{$\tau$\}                                                     \bigstrut\\
        BSL          &  \{$\tau_2$, $\tau_2/\tau_1$\}                                                           \bigstrut\\
        AdvInfoNCE  & \{$T_{\text{adv}}$, $E_{\text{adv}}$, $\text{lr}_{\text{adv}}$, $\tau$\}  \bigstrut\\
        \lossName\ (\textbf{IID})   &  \{$\text{lr}_{\text{margin}}$, $\gamma$ ,$\beta_0$\}                                                               \\
        \lossName\ (\textbf{OOD})  &  \{$\text{lr}_{\text{margin}}$, $\gamma$, $c_{\gamma}(\eta)$ ,$\beta_0$\}                                                               \bigstrut[b]\\
        \Xhline{1.3pt}
    \end{tabular}
        \label{tab:hyperparameters-searching}
\end{table}

\begin{table*}[htbp]
    \scriptsize
    \centering
    \caption{Optimal hyper-parameters of IID setting.}
    \begin{tabular}{c|l|ccc|ccc}
    \Xhline{1.2pt}
    \multirow{2}[4]{*}{\textbf{Model}} & \multicolumn{1}{c|}{\multirow{2}[4]{*}{\textbf{Loss}}} & \multicolumn{3}{c|}{\textbf{Gowalla}} & \multicolumn{3}{c}{\textbf{AmazonKitchen}} \bigstrut\\
    \cline{3-8}          &       & \textbf{lr} & \textbf{wd} & \textbf{others} & \textbf{lr} & \textbf{wd}  & \textbf{others} \bigstrut\\
    \Xhline{1.0pt}
    \multirow{8}[2]{*}{MF} 
    & MSE           & $10^{-3}$ & $10^{-7}$    &               & $10^{-4}$ & $10^{-2}$         &   \bigstrut[t]\\
    & BCE           & $10^{-3}$ & $10^{-3}$    &               & $10^{-3}$ & $10^{-2}$         &   \bigstrut[t]\\
    & BPR           & $10^{-4}$ & $10^{-3}$    &               & $10^{-3}$ & $10^{-4}$         &   \bigstrut[t]\\
    & CCL        & $10^{-2}$ & 0             & \{130, 0.80\}       & $10^{-1}$ & $0$                 &   \{9, 0.85\}      \\
    & LLPAUC            & $10^{-3}$ & 0      & \{0.80, 0.10\}       & $10^{-3}$ & $0$     &   \{0.70, 0.05\}      \\
    & SL           & $10^{-2}$ & 0           & \{0.09\}       & $10^{-2}$ & 0             &   \{0.20\}      \\
    & BSL           & $10^{-2}$ & 0          & \{0.08, 0.98\}       & $10^{-2}$ & 0       &   \{0.21, 0.95\}      \\
    & AdvInfoNCE    & $10^{-2}$ & $0$        & \{10, 20, $10^{-6}$, 0.09\}       & $10^{-2}$ & 0             &  \{15, 25, $10^{-5}$, 0.19\}     \\
    & DrRL        & $10^{-5}$ & $0$          & \{$10^{-5}$, 1.09, 0.85\}       & $10^{-3}$ & $0$             &   \{$10^{-4}$ 1.2, 0.85\}      \\
    \Xhline{1.0pt}
    \multirow{8}[2]{*}{LGCN} 
    & MSE           & $10^{-3}$ & $10^{-3}$ &                & $10^{-3}$ & $10^{-4}$                &    \bigstrut[t]\\
    & BCE           & $10^{-3}$ & $10^{-8}$ &                & $10^{-3}$ & $10^{-4}$                &     \bigstrut[t]\\
    & BPR           & $10^{-3}$ & $10^{-6}$ &             & $10^{-3}$ & $10^{-7}$                   &  \bigstrut[t]\\
    & CCL        & $10^{-3}$ & 0         &   \{160, 0.90\}       & $10^{-3}$ & $0$                  &   \{7, 0.85\}      \\
    & LLPAUC            & $10^{-3}$ & 0      & \{0.05, 0.10\}       & $10^{-4}$ & $0$               &   \{0.30, 0.05\}      \\
    & SL           & $10^{-1}$ & 0        & \{0.08\}       & $10^{-1}$ & 0                          &   \{0.25\}      \\
    & BSL           & $10^{-1}$ & 0        & \{0.08, 1.04\}       & $10^{-4}$ & 0                   &   \{0.23, 1.65\}      \\
    & AdvInfoNCE    & $10^{-2}$ & 0        & \{20, 5, $10^{-5}$, 0.08\}       & $10^{-4}$ & 0       &  \{15, 30, $10^{-5}$, 0.21\}     \\
    & DrRL        & $10^{-1}$ & $0$         & \{$10^{-4}$, 1.08, 0.90\}       & $10^{-4}$ & $0$     &   \{$10^{-4}$, 1.21, 0.85\}      \\
    \Xhline{1.0pt}
    \multirow{8}[2]{*}{XSimGCL} 
    & MSE           & $10^{-3}$ & $10^{-2}$ &    & $10^{-4}$ & $0$     & \bigstrut[t]\\
    & BCE           & $10^{-3}$ & $0$ &          & $10^{-3}$ & $0$     &    \bigstrut[t]\\
    & BPR           & $10^{-3}$ & $0$ &          & $10^{-3}$ & $0$     & \bigstrut[t]\\
    & CCL        & $10^{-2}$ & 0                 & \{160, 0.85\}       & $10^{-2}$ & $0$      &   \{8, 0.80\}      \\
    & LLPAUC            & $10^{-2}$ & 0          & \{0.05, 0.10\}       & $10^{-3}$ & $0$                &   \{0.50, 0.01\}      \\
    & SL           & $10^{-1}$ & 0         & \{0.07\}       & $10^{-2}$ & 0 &   \{0.18\}      \\
    & BSL           & $10^{-1}$ & 0         & \{0.07, 0.80\}       & $10^{-2}$ & 0 &   \{0.20, 1.22\}      \\
    & AdvInfoNCE    & $10^{-2}$ & 0         &   \{5, 1, $10^{-5}$, 0.08\}       & $10^{-2}$ & 0      &  \{10, 30, $10^{-6}$, 0.2\}     \\
    & DrRL        & $10^{-1}$ & $0$         &   \{$10^{-4}$, 1.09, 0.90\}       & $10^{-2}$ & $0$ &   \{$10^{-4}$, 1.24, 0.85\}      \\
    \Xhline{1.2pt}
    \multirow{2}[4]{*}{\textbf{Model}} & \multicolumn{1}{c|}{\multirow{2}[4]{*}{\textbf{Loss}}} & \multicolumn{3}{c|}{\textbf{AmazonElectroncis}} & \multicolumn{3}{c}{\textbf{AmazonBeauty}} \bigstrut\\
    \cline{3-8}          &       & \textbf{lr} & \textbf{wd} & \textbf{others} & \textbf{lr} & \textbf{wd} & \textbf{others} \bigstrut\\
    \Xhline{1.0pt}
    \multirow{9}[2]{*}{MF} 
    & MSE           & $10^{-3}$ & $10^{-2}$ &               & $10^{-2}$ & $10^{-2}$                    &   \bigstrut[t]\\
    & BCE           & $10^{-3}$ & $10^{-2}$ &               & $10^{-3}$ & $10^{-2}$                    &   \bigstrut[t]\\
    & BPR           & $10^{-3}$ & $10^{-3}$ &              & $10^{-3}$ & $10^{-5}$                     &   \bigstrut[t]\\
    & CCL        & $10^{-3}$ & 0         &    \{6, 0.90\}       & $10^{-4}$ & $0$                      &   \{9, 0.85\}      \\
    & LLPAUC            & $10^{-1}$ & 0         &    \{0.70, 0.10\}       & $10^{-2}$ & $0$            &   \{0.80, 0.01\}      \\
    & SL           & $10^{-1}$ & 0         &     \{0.26\}       & $10^{-1}$ & 0                        &   \{0.20\}      \\
    & BSL           & $10^{-1}$ & 0         &     \{0.25, 1.05\}       & $10^{-4}$ & 0                 &   \{0.20, 1.50\}      \\
    & AdvInfoNCE    & $10^{-1}$ & 0         &     \{10, 1, $10^{-4}$, 0.22\}       & $10^{-1}$ & 0     &  \{10, 30, $10^{-3}$, 0.23\}     \\
    & DrRL        & $10^{-1}$ & $0$         &     \{$10^{-5}$, 1.27, 0.70\}       & $10^{-4}$ & $0$    &   \{$10^{-5}$, 1.15, 0.90\}      \\
    \Xhline{1.0pt}
    \multirow{9}[2]{*}{LGCN} 
    & MSE           & $10^{-3}$ & $10^{-6}$ &                  & $10^{-2}$ & $10^{-2}$                     &   \bigstrut[t]\\
    & BCE           & $10^{-3}$ & $10^{-7}$ &                 & $10^{-3}$ & $10^{-3}$                      &    \bigstrut[t]\\
    & BPR           & $10^{-3}$ & $10^{-4}$ &                  & $10^{-3}$ & $10^{-7}$                     &   \bigstrut[t]\\
    & CCL        & $10^{-4}$ & 0            & \{6, 0.85\}       & $10^{-2}$ & $0$                          &   \{8, 0.85\}      \\
    & LLPAUC            & $10^{-3}$ & 0           & \{0.50, 0.05\}       & $10^{-3}$ & $0$                 &   \{0.80, 0.05\}      \\
    & SL           & $10^{-2}$ & 0          & \{0.25\}       & $10^{-2}$ & 0                               &   \{0.21\}      \\
    & BSL           & $10^{-3}$ & 0         & \{0.26, 1.40\}       & $10^{-1}$ & 0                         &   \{0.17, 1.45\}      \\
    & AdvInfoNCE    & $10^{-3}$ & 0         & \{10, 30, $10^{-4}$, 0.23\}       & $10^{-2}$ & 0            &  \{15, 30, $10^{-6}$, 0.23\}     \\
    & DrRL        & $10^{-1}$ & $0$         & \{$10^{-4}$, 1.21, 0.80\}       & $10^{-2}$ & $0$            &   \{$10^{-4}$, 1.17, 0.9\}      \\
    \Xhline{1.0pt}
    \multirow{9}[2]{*}{XSimGCL} 
    & MSE           & $10^{-3}$ & $10^{-7}$ &               & $10^{-3}$ & $10^{-2}$         &   \bigstrut[t]\\
    & BCE           & $10^{-3}$ & $0$ &           & $10^{-2}$ & $10^{-6}$       &   \bigstrut[t]\\
    & BPR           & $10^{-3}$ & $0$ &           & $10^{-3}$ & $0$        &   \bigstrut[t]\\
    & CCL        & $10^{-2}$ & 0                  & \{6, 0.80\}       & $10^{-2}$ & $0$                 &   \{9, 0.85\}      \\
    & LLPAUC            & $10^{-3}$ & 0           & \{0.50, 0.05\}       & $10^{-3}$ & $0$                &   \{0.50, 0.05\}      \\
    & SL           & $10^{-2}$ & 0                & \{0.23\}       & $10^{-1}$ & 0                  &   \{0.18\}      \\
    & BSL           & $10^{-1}$ & 0               & \{0.33, 1.35\}       & $10^{-1}$ & 0                &   \{0.17, 1.60\}      \\
    & AdvInfoNCE    & $10^{-2}$ & 0               & \{15, 20, $10^{-6}$, 0.3\}       & $10^{-2}$ & 0                 &  \{5, 15, $10^{-5}$, 0.21\}     \\
    & DrRL        & $10^{-2}$ & $0$               & \{$10^{-4}$, 1.19, 0.85\}       & $10^{-2}$ & $0$                &   \{$10^{-5}$, 1.16, 0.85\}      \\
    \Xhline{1.2pt}
    \end{tabular}
                \label{table:IID-Hyperparameter-official}
\end{table*}

\begin{table*}
    \scriptsize
    \centering
    \caption{Optimal hyperparameters of OOD setting.}
    \begin{tabular}{c|l|ccc|ccc}
    \Xhline{1.2pt}
    \multirow{2}[4]{*}{\textbf{Model}} & \multicolumn{1}{c|}{\multirow{2}[4]{*}{\textbf{Loss}}} & \multicolumn{3}{c|}{\textbf{amazonKitchen}} & \multicolumn{3}{c}{\textbf{amazonElectronics}} \bigstrut\\
    \cline{3-8}          &       & \textbf{lr} & \textbf{wd}  & \textbf{others} & \textbf{lr} & \textbf{wd}   & \textbf{others} \bigstrut\\
    \Xhline{1.0pt}
    \multirow{9}[2]{*}{MF} 
    & BPR           & $10^{-3}$ & $0$          &                        & $10^{-4}$ & $0$             &   \bigstrut[t]\\
    & CCL        & $10^{-2}$ & 0               & \{9, 0.90\}            & $10^{-2}$ & $0$             &   \{8, 0.80\}      \\
    & LLPAUC            & $10^{-3}$ & 0        & \{0.40, 0.05\}         & $10^{-3}$ & $0$             &   \{0.90, 0.10\}      \\
    & SL           & $10^{-1}$ & 0             & \{0.19\}               & $10^{-2}$ & 0               &   \{0.22\}      \\
    & BSL           & $10^{-3}$ & 0            & \{0.19, 1.35\}         & $10^{-2}$ & 0               &   \{0.18, 1.40\}      \\
    & AdvInfoNCE    & $10^{-3}$ & 0            & \{10, 5, $10^{-5}$, 0.20\}         & $10^{-2}$ & 0              &  \{10, 15, $10^{-6}$, 0.25\}     \\
    & DrRL        & $10^{-2}$ & $0$            & \{$10^{-4}$, 2.50, 5, 0.85\}       & $10^{-1}$ & $0$            &   \{$10^{-5}$, 1.16, 1.25, 0.90\}      \\
    \Xhline{1.2pt}
    \end{tabular}
        \label{tables:OOD-hyperparameter}
\end{table*}

\end{document}